\documentclass[]{aastex631}

\usepackage{multirow}
\shorttitle{Variation on Volumetric Evolution of CMEs}
\shortauthors{Majumdar et al.}
\graphicspath{{./}{figures/}}
\begin{document}

\title{On the Variation of Volumetric Evolution of CMEs from Inner to Outer Corona}

\author[0000-0002-6553-3807]{Satabdwa Majumdar}
\affiliation{Indian Institute of Astrophysics, 2nd Block, Koramangala, Bangalore, 560034, India}

\author[0000-0001-8504-2725]{Ritesh Patel}
\affiliation{Indian Institute of Astrophysics, 2nd Block, Koramangala, Bangalore, 560034, India}
\affiliation{Aryabhatta Research Institute of Observational Sciences, Nainital, 263001, India }

\author[0000-0002-6954-2276]{Vaibhav Pant}
\affiliation{Aryabhatta Research Institute of Observational Sciences, Nainital, 263001, India }



\begin{abstract}

 Some of the major challenges faced in understanding the early evolution of Coronal Mass Ejections (CMEs) are due to limited observations in the inner corona ($<\,3$ R$_{\odot}$) and the plane of sky measurements. In this work, we have thus extended the application of the Graduated Cylindrical Shell (GCS) model to the inner coronal observations from the ground--based coronagraph K--Cor of the Mauna Loa Solar Observatory (MLSO) along with the pair of observations from COR--1 onboard the Solar Terrestrial Relations Observatory (STEREO).  We study the rapid initial acceleration and width expansion phase of 5 CMEs in white-light in the lower heights. We also study the evolution of the modelled volume of these CMEs in inner corona and report for the first time, a power law dependence of the CME volume with distance from the Sun. We further find the volume of ellipsoidal leading front and the conical legs follow different power laws, thus indicating differential volume expansion through a CME.  The study also reveals two distinct power laws for the total volume evolution of CMEs in the inner and outer corona, thus suggesting different expansion mechanisms at these different heights. These results besides aiding our current understanding on CME evolution, will also provide better constraints to CME initiation and propagation models. Also, since the loss of STEREO-B (and hence COR--1B data) from 2016, this modified GCS model presented here will still enable stereoscopy in the inner corona for the 3D study of CMEs in white-light. 

\end{abstract}
%We present here a proof of concept of the application of this modified GCS model to K--Cor data for the first time and thus study the 3D evolution of CMEs uniquely in white-light (WL) observations in the inner corona.



\section{Introduction} \label{sec:intro}

One of the most fascinating and intriguing phenomena occurring in the Sun's corona are the Coronal Mass Ejections (CMEs), which involve large scale release of magnetised plasma outwards into the heliosphere. They are most generally defined as discrete, bright, white-light features propagating outwards in the coronagraph field of view (FOV) \citep{hundhausen_1984}. They largely vary in their shapes and appearances, and are known to show a wide range in their kinematic properties \citep[for a review, see][]{liv_rev_2012}. Apart from that, CMEs are also the major drivers of space weather and the ones travelling towards Earth can have a severe impact on it by creating geomagnetic storms that can pose a threat to our several technological advancements and life as a whole \citep{schwenn2005,gosling_1993}. This demands better preparation for such a plausible event of chance, and hence as a prerequisite, a very good understanding of their kinematics. 

It has been known that CME kinematics is an outcome of the interplay of three forces, namely the Lorentz force, the gravitational force and the viscous drag force, the latter arising due to the interaction with the ambient solar wind \citep{liv_rev_2012}. The outcome of this interplay of forces is reflected into a three--phase kinematic profile, with an initial gradual rise phase, followed by an impulsive phase, and then a residual propagation phase \citep{Zhang_2001,Zhang_2004}. The initial rise phase is marked by a very weakly accelerated motion \citep{cheng_2020}, while the later residual phase is seen as propagation with almost constant or decreasing speed \citep[see][]{nat_2000}. The main impulsive acceleration phase however, is qualitatively very different from the initial slow rise phase and involves a rapid increase in acceleration in a short interval of time, that shoots the CMEs to high velocities \citep[e.g.][]{bein_2011,cheng_2020, patel2020}. Earlier studies have suggested that this main acceleration phase occurs at the lower coronal heights and hence might not be always captured using traditional white-light coronagraphic observations \citep{gallagher_2003,Temmer_2008,majumdar_2021}. Earlier attempts at measurements of this main acceleration phase have been reported by several studies in the past. In most of these works, the method either relied on measurements on the plane of the sky, thus introducing discrepancies due to projection effects \citep[e.g.][]{stCyr_1999,Zhang_2006,Balmaceda_2018}, or involve combining white-light with Extreme Ultraviolet (EUV) data for tracking a CME \citep[e.g.][]{vrsnak_2007,bein_2011}, where whether the same features are observed in emission lines and in white-light are still debatable \citep[see][]{song_2019}. Now, although we do have now an understanding of the impact of drag force on the kinematics \citep[][and references therein]{sachdeva_2015}, the impact of Lorentz force still eludes a clear understanding.  Recently \cite{majumdar_2020} used the Graduated Cylindrical Shell (GCS) model \citep[developed by][]{Thernisien_2006,thernesien_2009,thernesien_2011} to study the 3D evolution of CMEs in the inner and outer corona, and reported that the true height till which the imprint of Lorentz force remains dominant lies in the range $2.5-3$ R$_{\odot}$, thus further indicating the importance of inner corona observations.

CMEs, apart from radial propagation, also show lateral expansion of their angular width \citep[see][]{Kay_2015} until a certain critical height, after which they propagate with almost constant width \citep[e.g.][]{Moore_2007,zhao_2010}. The usual method of width estimation involves the projected angular span between the position angles of the two extreme flanks of the CME \citep{zhao_2010}, but such estimation suffers from a lot of projection effects. In this regard, \cite{cremades_2020} used the GCS model to study the axial and lateral width expansion of CMEs by combining white-light and EUV observations. Also, \cite{majumdar_2020}, using the GCS model, reported on the observational evidence that the angular width expansion and the impulsive accelerations are just  manifestations of the same Lorentz force, as conjectured earlier by \cite{Subramanian_2014,Suryanarayana_2019}.  In this regard, it was further reported that the evolution and width expansion of CMEs is non self-similar in the inner corona \citep[][]{cremades_2020}, while it is self-similar in the outer corona \citep[][]{Subramanian_2014}. It is also worth noting that the distribution of angular widths of slow and fast CMEs from different source regions have been known to follow different power law profiles, thus indicating the possibility of different generation mechanisms \citep[as reported recently by][]{pant_2021}. Thus, a study of the evolution of CME volume (which is influenced by the width expansion of CMEs) would shed more light on this aspect of CME evolution. In this regard, \cite{Holzknecht_2018} used the GCS model to estimate the volume of a CME. Later this treatment was also used by \cite{Temmer_2021} to study the density evolution of CMEs with distance from the Sun, but both these works reported on results in the outer corona and the heliosphere, and thus we do not have a good understanding of the evolution of total volume in the inner corona.

A major challenge in the understanding of early CME kinematics in the inner corona has been due to limited observational white-light data below 3 R$_{\odot}$ and projection effects. Several techniques have been developed to address the later issue \citep[see][]{Mierla_2008,thernesien_2009,Joshi_2011,Hutton_2017}, but the implementation of such techniques to inner corona has been limited. To address these shortcomings, we extend the implementation of the GCS model to the inner corona observations from the ground--based coronagraph K--Cor of the Mauna Loa Solar Observatory (MLSO) which offers a FOV of $1.05-3$ R$_{\odot}$. This will enable us to capture the initial impulsive phase of the CMEs uniquely in white light observations. Using this extended GCS model, we thus study the early 3D evolution of 5 CMEs by studying their kinematic profiles, widths and volume evolution as they propagate from the inner to the outer corona. We outline the data source and working method in Section~\ref{sec2}, followed by our results in Section~\ref{sec3}, and we present our main conclusions and discussions in Section~\ref{sec4}.   

\section{Data and Method} \label{sec2}

\subsection{Data Source and Data Preparation}

The data used in this work are taken from coronagraphs COR--1 (FOV of 1.5--4 R$_\odot$), COR--2  (FOV of 2.5--15 R$_\odot$) and Extreme UltraViolet Imager (EUVI) of the the Sun Earth Connection Coronal and Heliospheric Investigation package \citep[SECCHI][]{secchi} on-board the twin spacecraft Solar Terrestrial Relations Observatory \citep[STEREO;][]{stereo}, the K-Cor (DOI: 10.5065/D69G5JV8) ground based coronagraph (FOV of 1.05--3 R$_\odot$) of the MLSO and the data from Large Angle Spectroscopic COronagraph \citep[LASCO][]{Brueckner_1995}  (FOV of 2.2--30 R$_\odot$). Level 0.5 data of EUVI, COR--1 and COR--2 was reduced to Level 1.0 using the $secchi\_prep.pro$ routine in IDL. For the K-Cor data, we used the 2 min cadence Level 2.0 data processed through the Normalized Radially Graded Filter \citep[NRGF;][]{morgan_2006,nrgf}, and for LASCO, we used level 1 data (corrected for instrumental effects, solar North and calibrated to physical units of brightness). Finally, base difference images were created for K--Cor, COR--1, COR--2 and LASCO by subtracting a pre--event image from successive images of the event thereafter.

\subsection{Event Selection} \label{event_selection}
Since this work involves combining data from the COR--1, COR--2 coronagraphs on STEREO, LASCO coronagraphs (FOV of 1.5--4 R$_\odot$) on SOHO and K--Cor of MLSO, only those events could be selected which were simultaneously observed by these instruments. It should be noted here that K-Cor and LASCO are not simultaneously used by the GCS model, but rather LASCO is replaced by K-Cor for the lower coronal heights. Now, K--Cor being a ground--based coronagraph, only the day time observations are available  (approximately 17:30 UT to 02:30 UT), and this largely restricted the event selection. Also, those CMEs were selected which had a distinct leading edge in the FOV of the above coronagraphs, thus assuring unambiguous tracking in the successive frames. Since K--Cor views the solar corona through the Earth's atmosphere, the data is affected by weather conditions. Also, the identified CMEs tend to be fainter in K--Cor as compared to COR--1, thus rendering tracking more challenging. This can be due to the bright sky background leading to a low signal or due to the fact that CMEs tend to gather mass at these low heights \citep{thompson_2017}. Based on the above criteria, 5 CMEs were selected from the K-Cor catalogue that occurred between 2014 February and 2016 January.

%\begin{figure*}[h]
%\gridline{%\fig{20140212_222250.png}{0.25\textwidth}{(e) 22:22 UT}
 %         \fig{20140212_222451.png}{0.25\textwidth}{(a) 2014--02--12 (22:24 UT)}
  %        \fig{20140212_222652.png}{0.25\textwidth}{(b) 2014--02--12 (22:26 UT)}
   %       \fig{20140212_222854.png}{0.25\textwidth}{(c) 2014--02--12 (22:28 UT)}
    %     }

%\caption{The evolution of the CME in the inner corona in the K-Cor FOV for the CME that occurred on \textbf{\textcolor{red}{ February 12 2014}}.}
%\label{kcorviews}
%\end{figure*}

\subsection{The GCS fitting to STEREO and K--Cor data} \label{fit_algo}

The GCS model was developed to fit a synthetic flux--rope to a pair of coronagraph images taken from the two different vantage points offered by the positions of STEREO-A/B. A provision is also made for including observations from the LASCO coronagraphs as a third vantage point. To study the evolution of CMEs from inner corona, in this work, we first extended the model further to include observations of the inner corona from K-Cor of MLSO as a third vantage point (in the Sun--Earth line), as the FOV of K--Cor will largely aid in understanding the early evolution of CMEs. Since the header structure of K--Cor data is different than that of LASCO, the primary codes that generate the synthetic flux--rope, namely $rtsccguicloud.pro$ and $rtcloud.pro$ needed to be modified. Hence a similar block of code (as was present for LASCO) was developed for the K-Cor observations by introducing relevant keywords for K--Cor data corresponding to the keywords for LASCO data to the above procedures. This was added with a condition that simultaneously, either K--Cor or LASCO observations are to be present along with the STEREO observations. Thanks to the overlapping FOVs, K--Cor observations were combined with COR--1, and LASCO with COR--2 observations, thus ensuring a three vantage point tracking throughout. The novelty of this work also lies in the fact that despite the unavailability of STEREO--B observations after 2016, we can still perform stereoscopy in the inner corona by combining data from COR--1A and K--Cor with the help of this extended GCS model. In the following steps we outline the procedure of fitting carried out in this work:

Step 1 -- A pair of COR--1 images and a K--Cor image (taken at almost the same time) are selected where the CME front is well developed in all three images.

Step 2 -- The fitting procedure is then followed as outlined in \cite{thernesien_2009,majumdar_2020}.

Step 2 -- The above two steps are repeated for the successive images in which the CME front was well developed in both the K--Cor and COR--1 FOVs.

Step 3 -- Since the time of appearance of CME in K-Cor FOV might be different than the same in COR-1 FOV, a K-Cor image is then selected for which the CME front is first observed. Since 3 vantage point observations are not available for this height, some of the model parameters are fixed, while the height, half--angle and aspect--ratio are re--adjusted as the model is fitted to the K-Cor images. 

Step 4 -- Finally, the model is fitted to LASCO and the pair of COR--2 images to capture the evolution in the outer corona. The uncertainty in fitting is determined in a similar way as mentioned in \cite{thernesien_2009,majumdar_2020}.

\begin{figure*}[h]
\gridline{%\fig{cme_kcor_20140212.png}{0.25\textwidth}{(a) CME view in K-Cor}
          \fig{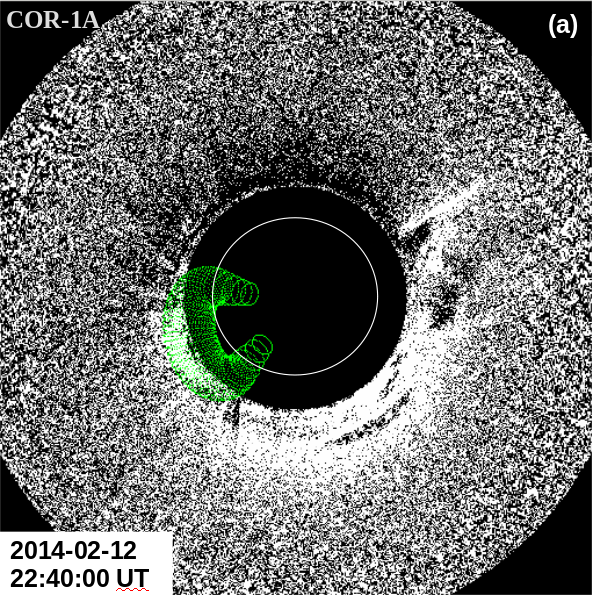}{0.25\textwidth}{} 
          \fig{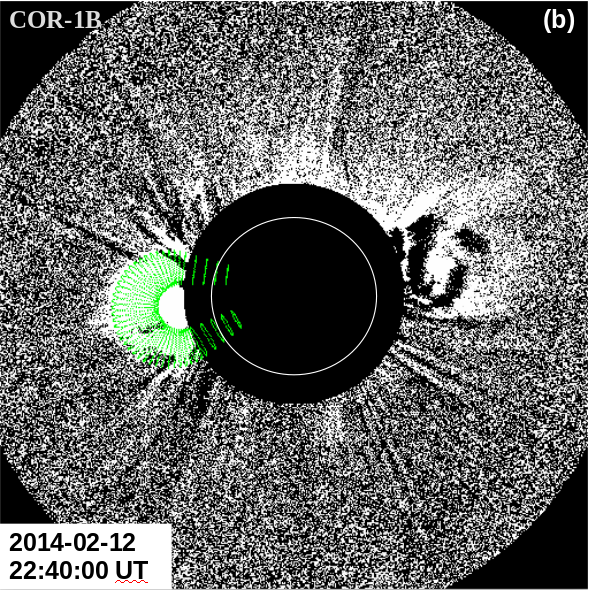}{0.25\textwidth}{}
          \fig{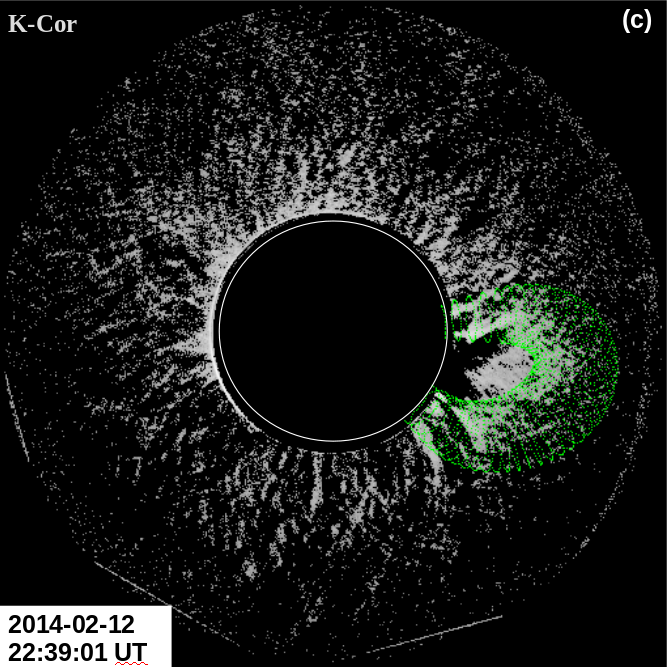}{0.25\textwidth}{}
         }
         \vspace{-0.048\textwidth}
\gridline{%\fig{cme_kcor_20140614.png}{0.25\textwidth}{(e) CME view in K-Cor}
          \fig{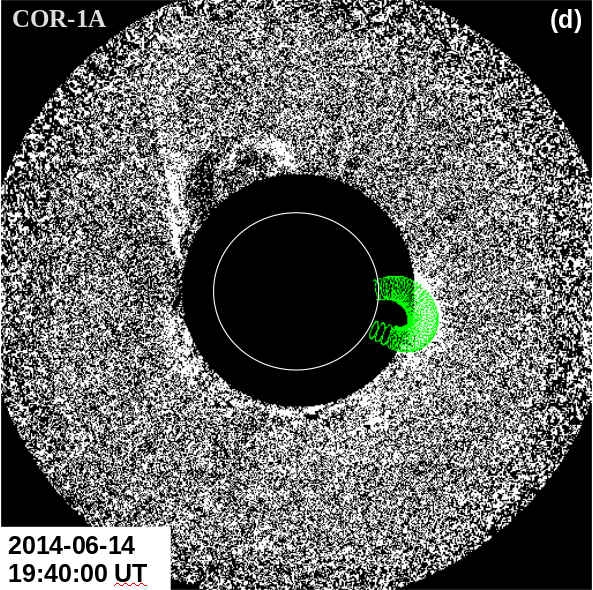}{0.25\textwidth}{}
          \fig{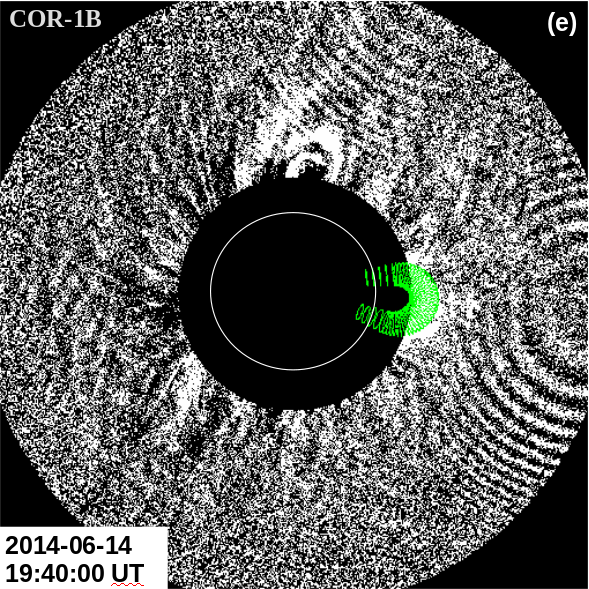}{0.25\textwidth}{}
          \fig{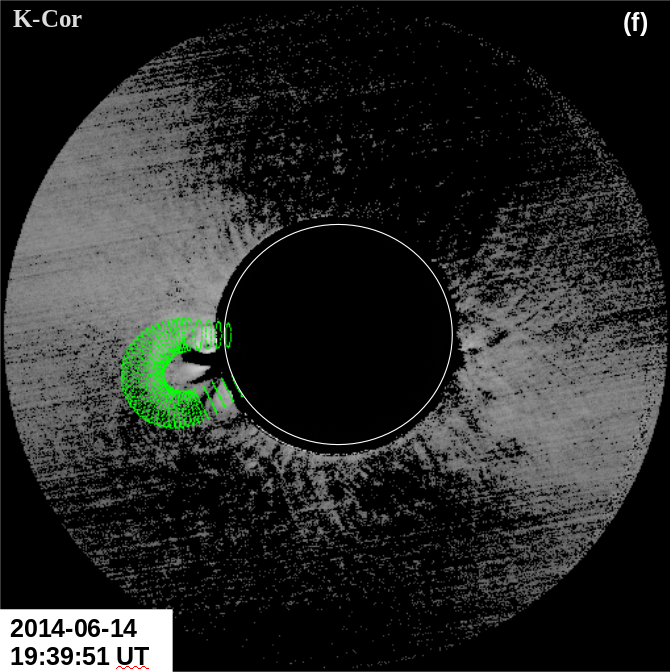}{0.25\textwidth}{}
         }    
         \vspace{-0.048\textwidth}
\gridline{%\fig{cme_kcor_20140614.png}{0.25\textwidth}{(e) CME view in K-Cor}
          \fig{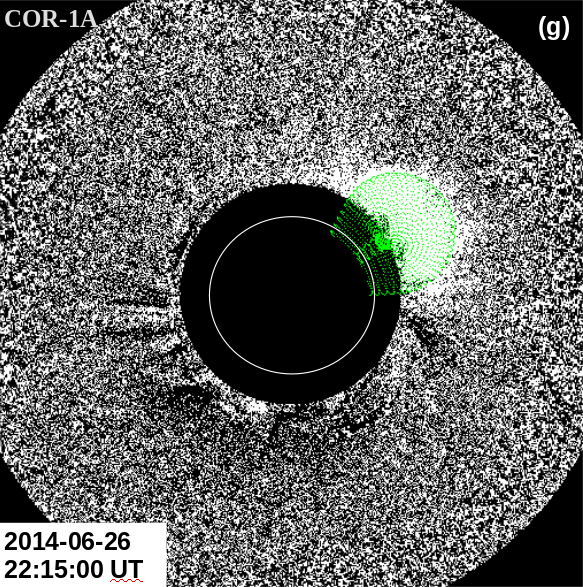}{0.25\textwidth}{}
          \fig{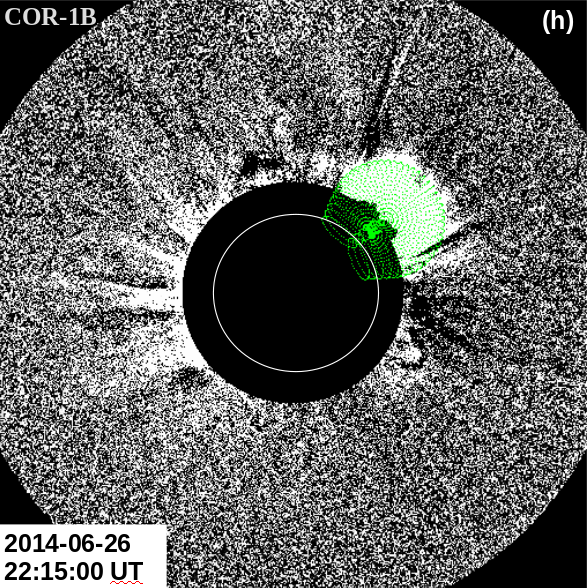}{0.25\textwidth}{}
          \fig{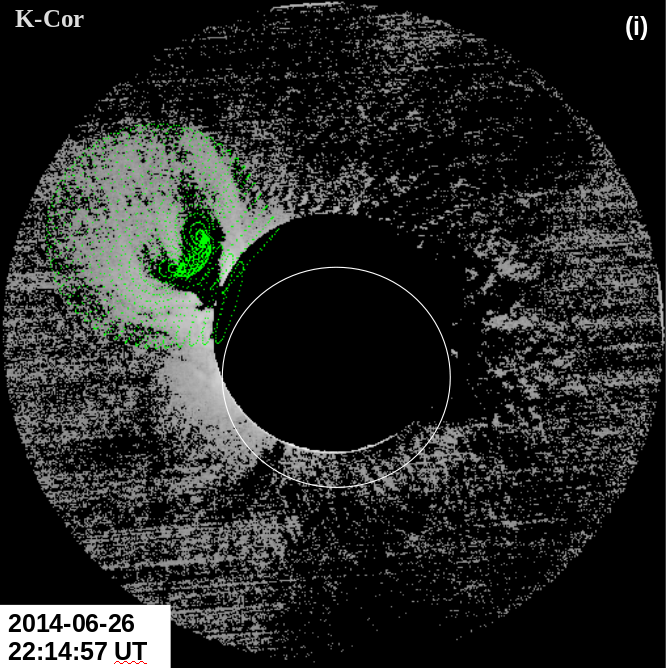}{0.25\textwidth}{}
         }   
         \vspace{-0.048\textwidth}
\gridline{%\fig{cme_kcor_20140614.png}{0.25\textwidth}{(e) CME view in K-Cor}
          \fig{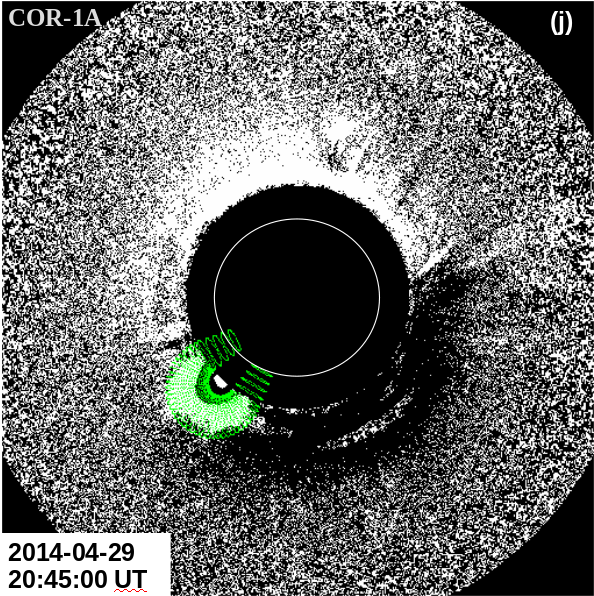}{0.25\textwidth}{}
          \fig{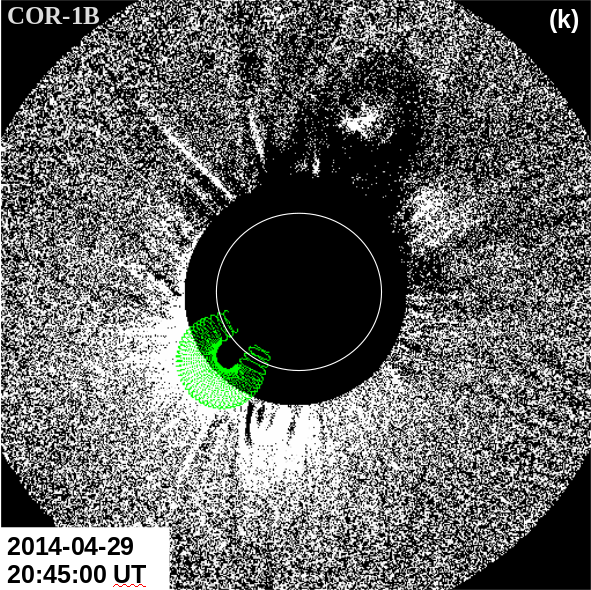}{0.25\textwidth}{}
          \fig{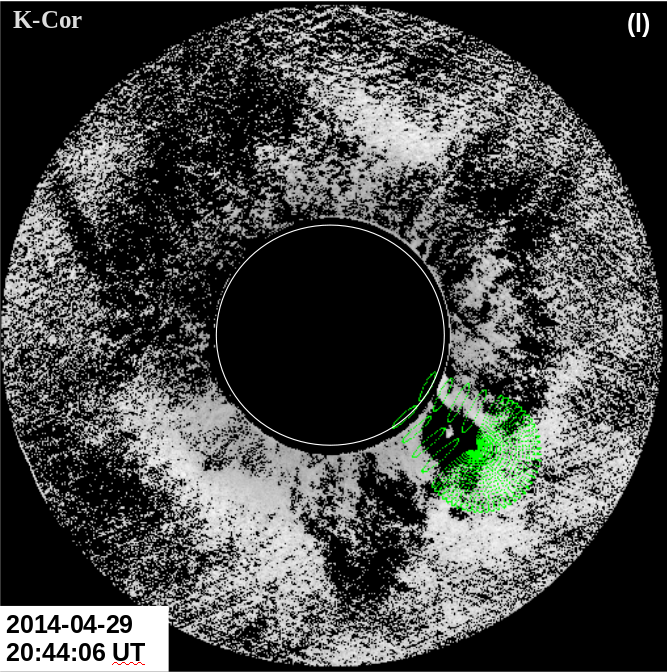}{0.25\textwidth}{}
         }    
         \vspace{-0.048\textwidth}
%\gridline{%\fig{cme_kcor_20140614.png}{0.25\textwidth}{(e) CME view in K-Cor}
 %         \fig{cor1a_20140614.png}{0.25\textwidth}{(d) COR-1A}
  %        \fig{cor1b_20140614.png}{0.25\textwidth}{(e) COR-1B}
   %       \fig{kcor_20140614.png}{0.25\textwidth}{(f) K-Cor}
    %     }   
\gridline{\fig{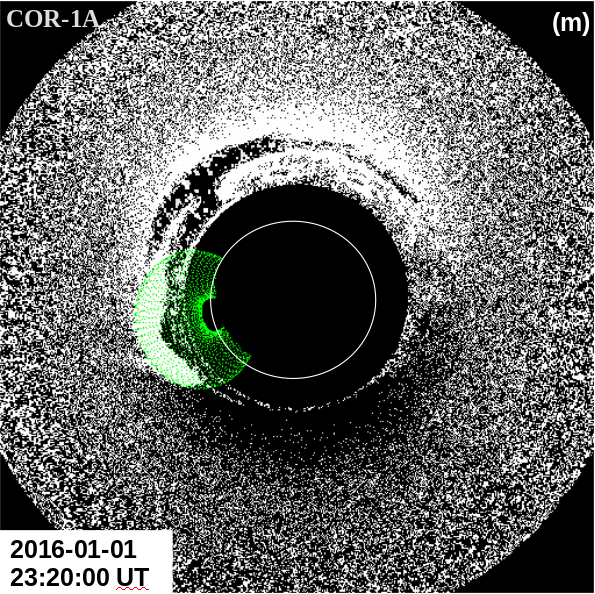}{0.25\textwidth}{}
          \fig{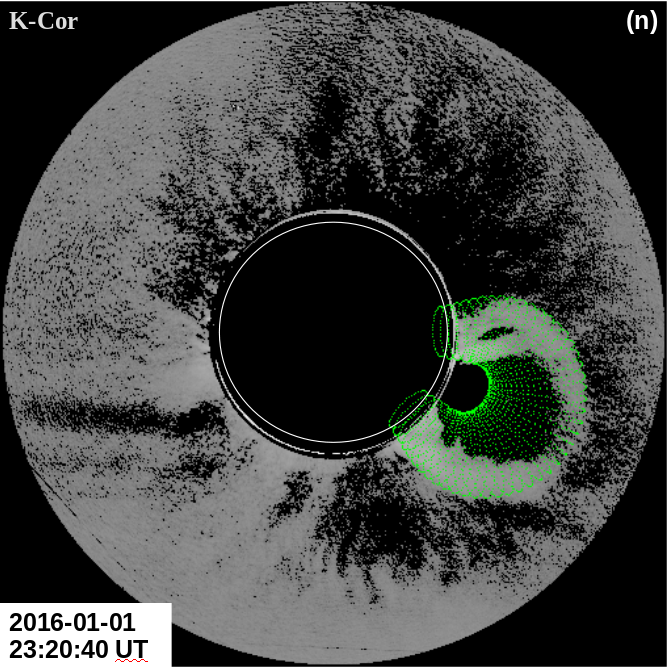}{0.25\textwidth}{}
         }   
         \vspace{-0.048\textwidth}
\caption{The fitting of the GCS flux-rope to K--Cor and the pair of COR--1 images for the 5 CMEs studied in this work.}
 \label{gcs_fits}
\end{figure*}

Examples of the GCS fitting to K--Cor and COR--1 images are shown in Figure~\ref{gcs_fits} and a summary of the fitted parameters are given in Table~\ref{table}. Panels (g) and (h) of Figure~\ref{gcs_fits} further reflect the significance of this extended GCS model, for the study of 3D kinematics in inner corona, despite the unavailability of COR-1B data. 

%After the model is fitted, the time evolution of the parameters are recorded for further analysis.

%It is worthwhile to note here that owing to the difference in the starting FOVs of K--Cor and COR--1, the CME leading front reaches almost the edge of K--Cor FOV (often becomes diffused), by the time it fully appears in the COR--1 FOV, which further restricted event selections and posed challenges to the GCS fitting.

\begin{table}[h]
    \centering
    \begin{tabular}{c|c|c|c|c|c|c|c}
    \hline
        Date & Time & Longitude ($\phi$) & Latitude ($\theta$) & Tilt angle ($\gamma$) & Height ($h$) & Aspet ratio ($k$) & Half-angle ($\alpha$)  \\
         & (hh:mm:ss) (UT) & (deg) & (deg) & (deg) & (R$_{\odot}$) &  & (deg) \\
         \hline
         \hline
        2014 February 12 & 22:40:00 & 102 & -10 & -45 & 2.79 & 0.16 & 24 \\
        2014 June 14 & 19:45:00 & 84 & -13 & 64 & 2.29 & 0.14 & 23 \\
        2014 June 26 & 22:15:00 & 290 & 29 & -68 & 2.86 & 0.36 & 13 \\
         2014 April 29 & 20:45:00 & 142 & -39 & -81 & 2.29 & 0.16 & 13 \\
        2016 January 1 & 23:20:00 & 330 & -22 & 83 & 2.47 & 0.22 & 22 \\
        \hline
    \end{tabular}
    \caption{The GCS model parameters fitted to the CMEs are tabulated. The Time is the time of observation, $\phi$, and $\theta$ are the longitude and latitude of the CME, the tilt-angle ($\gamma$) is the angle between the axis of symmetry of the CME and the solar equator, $h$ is the height of the leading front, aspect-ratio ($k$) is the ratio of the minor to the major CME radius and $\alpha$ is the half-angle between the legs of the CME.}
    \label{table}
\end{table}

\section{Results} \label{sec3}

\subsection{Improvement in the understanding of early CME kinematics}

\begin{figure*}[h]
\gridline{\fig{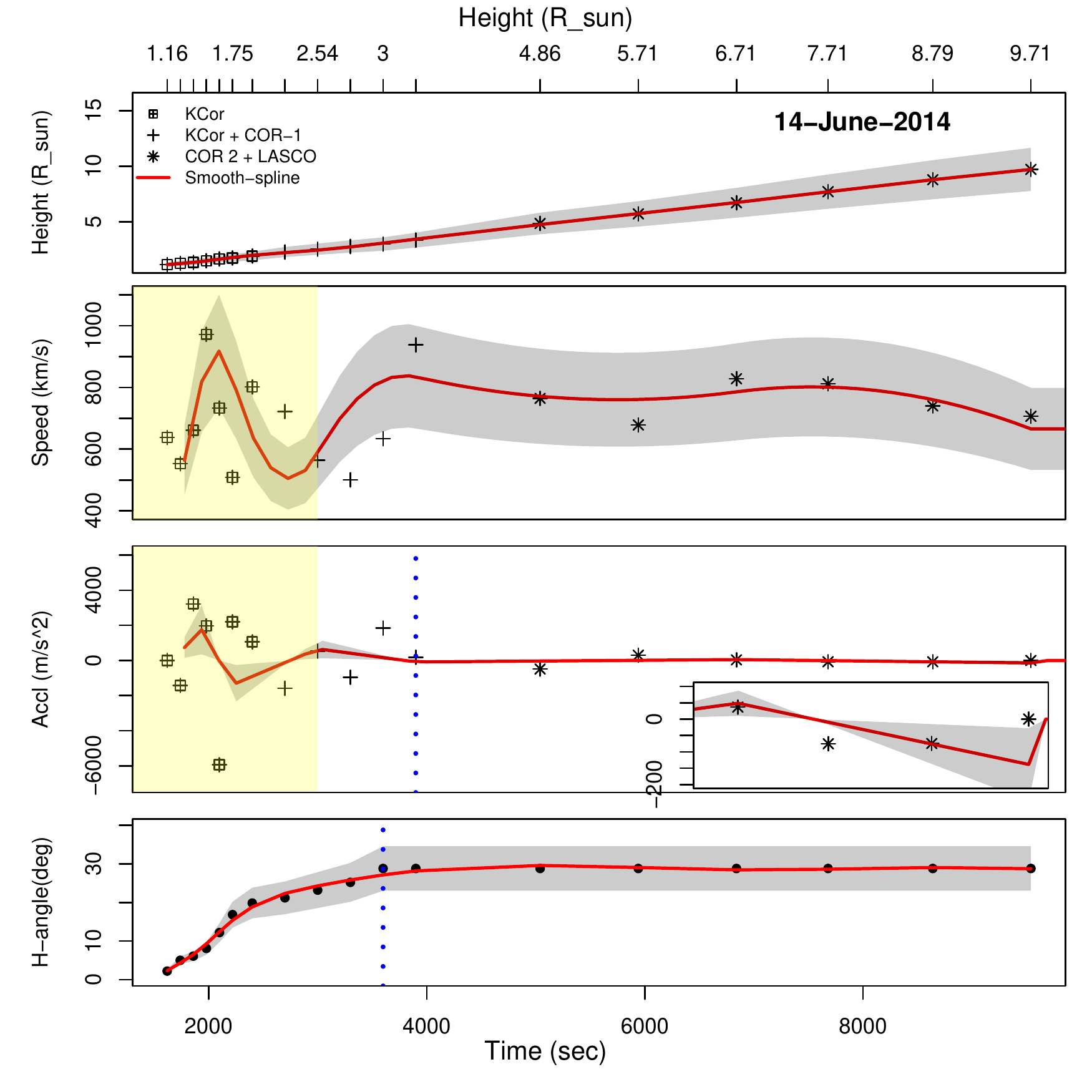}{0.45\textwidth}{(a) CME on June 14 2014}
          \fig{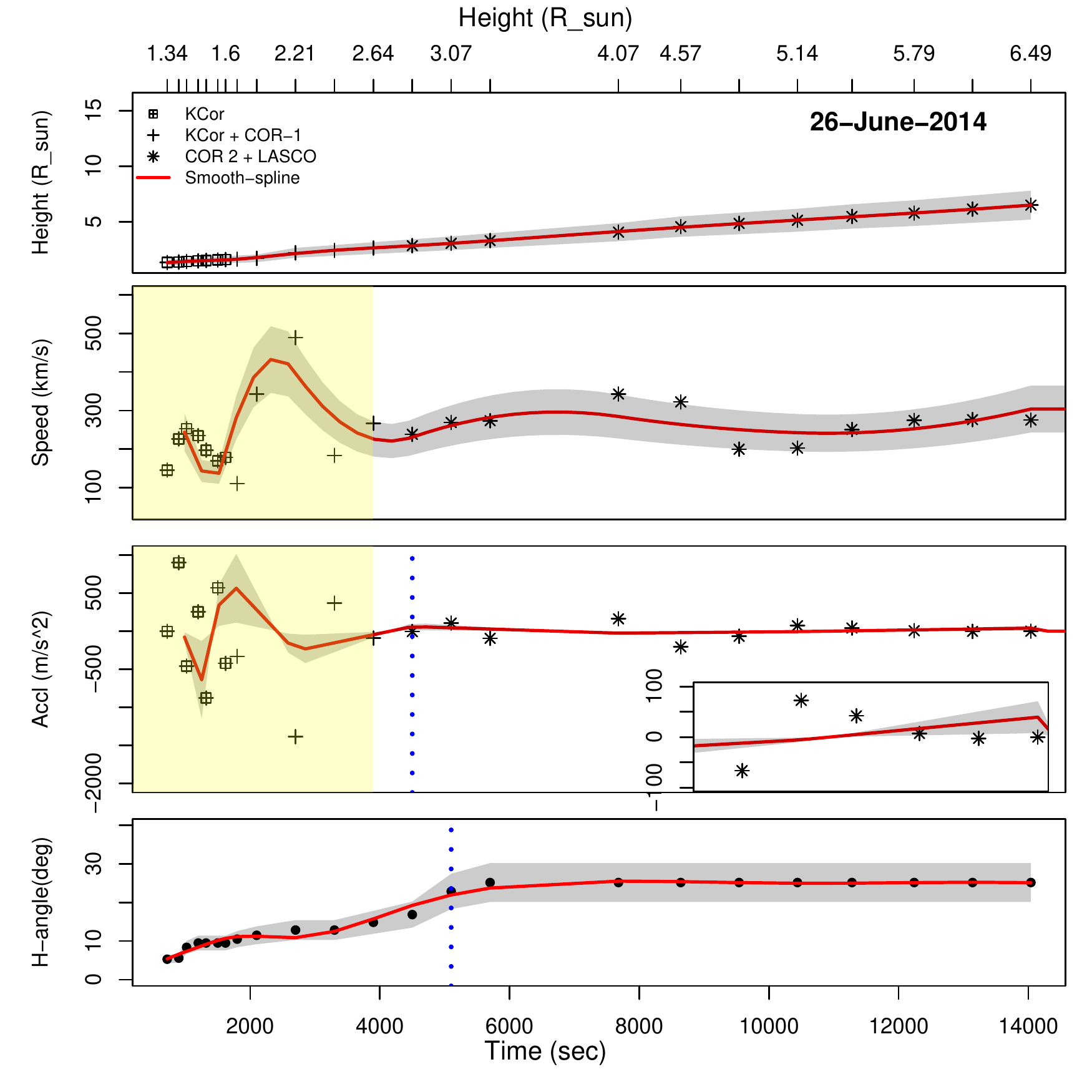}{0.45\textwidth}{(b) CME on June 26 2014}
          }
\caption{The complete 3D kinematic profiles of two out of five impulsive CMEs are shown as representative examples. The height-time data is fitted with a cubic smooth spline (shown in solid red line). The speed and acceleration plots are obtained by taking first and second order numerical derivatives of the height--time plot. The grey shaded region corresponds to the uncertainty in the fitted model parameters. The impulsive phase is highlighted in the second and third panels in yellow. An inset with a zoom into the residual acceleration phase is provided in the bottom right hand corner of the acceleration plots. Please note that the time axis of the zoomed insets is overlapped with the common time axis shown at the bottom. In the bottom, the evolution of the half--angle ($\alpha$) parameter is plotted.}
\label{kinem_profiles}
\end{figure*}

It should be noted here that although while selecting the events, no such pre-selection criteria was kept on the CMEs to be impulsive, yet it turned out that all the five CMEs studied, showed the impulsive phase. In Figure~\ref{kinem_profiles}(a) and (b), we plot the 3D kinematic profiles of the CMEs that occurred on June 14, 2014, and June 26, 2014. On the top we plot the height--time data fitted with a cubic smooth spline (in red), followed by the speed and acceleration profiles (derived by taking the first and second order numerical derivatives of the height-time data) in the second and third panels. The overall fitting procedure and the estimation of speed and acceleration is the same as reported in \cite{majumdar_2020}. It should be noted here that the average uncertainty in the fitting of the GCS model was found to be 20 percent, and we did not find any appreciable change in latitude/longitude of these events beyond their uncertainties. However, it is worthwhile to note that a change in latitude/longitude will influence the height measurements, and hence for events which show considerable deflections, these considerations should be taken into account in future in estimating the uncertainty region in the absolute lower heights in the height-time profiles. We also plot the variation of the half--angle parameter ($\alpha$) in the bottom panel. In the third panel, an inset with a zoomed in plot of the residual acceleration phase is also provided in the right hand bottom corner. Please note that the time axis of the zoomed insets is overlapped with the common time axis shown at the bottom. We find that with the aid of the observations from K--Cor, it was possible to capture the initial impulsive acceleration phase of the CMEs uniquely in the white-light data, thus escaping the need for combining EUV observations with white-light observations for capturing the same, as was reported earlier in \cite{bein_2011} (while the initial gradual rise phase seems to have been already  got over by the time the CMEs reached the K-Cor FOV). It is worthwhile to point out that this was not possible in \cite{majumdar_2020, cremades_2020}, as for a number of events, the impulsive acceleration phase was already over by the time the CME entered the COR--1 FOV, leading to an underestimate of the true acceleration, magnitude and duration. Please note that in Figure~\ref{kinem_profiles}, we show the kinematic profiles as representative examples of the two of the five impulsive CMEs studied, so as to demonstrate the capturing of the impulsive phase by only using white light observations Further, as K--Cor offers better cadence than COR--1 (in our case, we have used 2 minutes cadence data), it helps in better tracking of the CME in the lower heights. Nonetheless, it must be noted that during the tracking of CME in the K-Cor and COR-1 overlapping FOV, the fitted times will be limited by the cadence of COR-1. In this regard, we would like to point out that although K--Cor data offers a better cadence of fifteen seconds, the CME front in them was fainter and tracking it was difficult. It should also be noted that, sometimes the leading edge in the K-Cor image gets diluted in the higher heights of its FOV. Now, although, this would introduce an uncertainty in the measured height, yet it's worth noting that the application of the GCS model leads to the tracking of a certain front of the CME (in this case the leading front) and not a certain point on the leading front. Thus, in such cases, the other view points from COR-1, where the CME leading front is better visible helps in tracking the CME through those heights, while we use K-Cor observations to track the CME in lower heights (as mentioned in Section~\ref{fit_algo}), where the leading front is better visible. The blue vertical dotted lines in the acceleration and half-angle evolution plots denote the time (and height) at which the impulsive acceleration ceases and the half-angle becomes constant respectively. For the events studied, these heights happen to lie in the range of 2.5--3 R$_{\odot}$ \citep[consistent with][]{cremades_2020,majumdar_2020}.
% , thus supporting the earlier results of \cite{majumdar_2020}.
%The average speed and acceleration of the CME on June 14, 2014 was found to be 709 Kms$^{-1}$ and 43 ms$^{-2}$ and for the CME on June 26, 2014 was found to be 250 Kms$^{-1}$ and 20 ms$^{-2}$ respectively.

\subsection{Insights on width expansion of CMEs}

\begin{figure}
\gridline{\fig{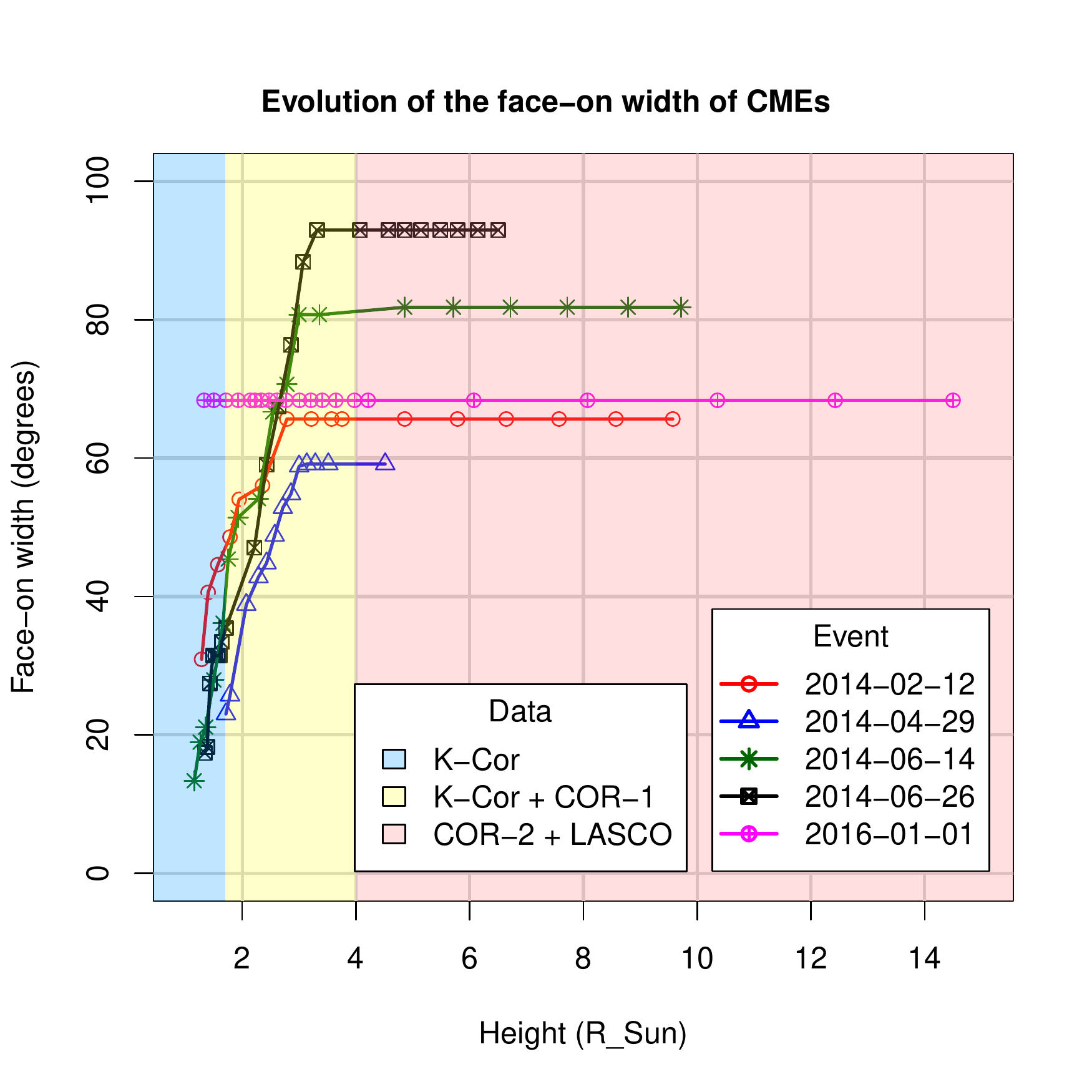}{0.45\textwidth}{(a) Face-on width evolution}
          \fig{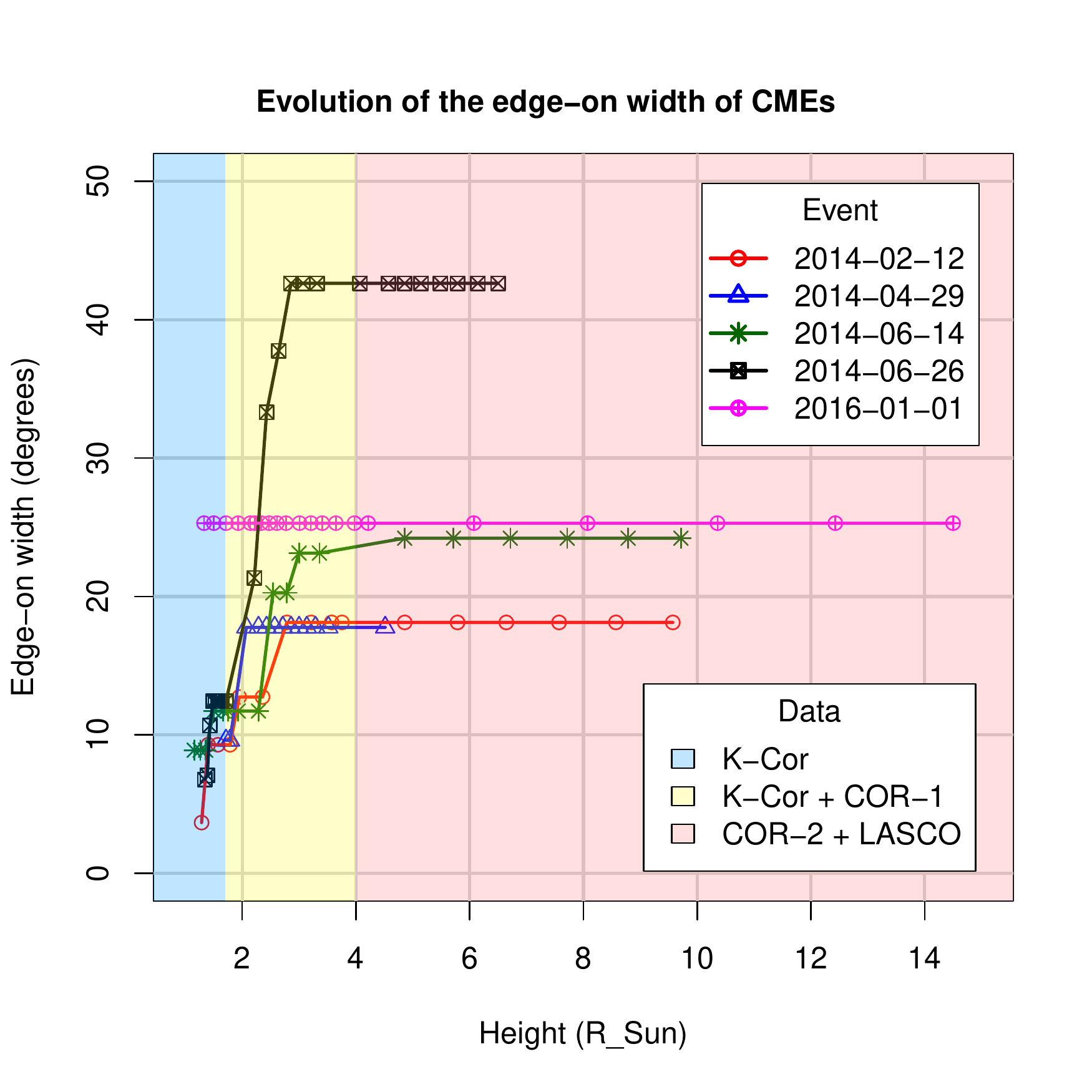}{0.45\textwidth}{(b) Edge-one width evolution}
          }  
\caption{The evolution of the modelled (a) face-on and (b) edge-on width of CMEs in the inner corona. Different regions of the plot are highlighted according to the data used.}
    \label{width_profile}
\end{figure}

Use of three vantage point observations helped in better constraining the GCS parameters (nevertheless it should be noted that for the CME in 2016, only two vantage points were available, and for the heights below COR-1 FOV, only K-Cor observations are used). Multiple vantage point observations have shown that the width of a CME can be seen in two broad perspectives. CMEs tend to expand along the direction of their main axis giving their axial width, and in the direction perpendicular to it giving their lateral width \citep{cabello_2016} which corresponds to the face--on (FO) and edge--on (EO) CME widths as presented in \cite{thernesien_2009}. Thus, instead of just studying the evolution of the half-angle parameter as a proxy of studying the width expansion, we use the half--angle ($\alpha$) and the aspect--ratio ($k$
) to calculate the FO and EO widths of the CMEs studied. This was possible once the GCS parameters for the CMEs were fixed by the three above-mentioned vantage points which were back-traced in the K-Cor FOV to heights of $\approx$1.1 R$_\odot$. From Table 1 of \cite{thernesien_2011}, the FO width ($f_w$) is related as,

\begin{equation}
    f_w \,=\, 2\,(\alpha + Sin^{-1}k)
\end{equation} \label{eqn1}

and the EO width ($e_w$) is related as,

\begin{equation}
    e_w \,=\, 2\,Sin^{-1}k.
\end{equation} \label{eqn2}

In Figure~\ref{width_profile}(a) and (b) we plot the variation of the FO and EO widths of the CMEs with height. We find that initially, until 3 R$_{\odot}$, both the FO and EO widths increase rapidly with height and then saturates, thus implying that in these lower heights, CMEs expand rapidly in both the axial and lateral directions. A similar behaviour was also reported by \cite{cremades_2020}, but it should be noted that they combined EUV and white-light observations to arrive at this conclusion, while our conclusions are based on using only white-light data uniquely. In this context, it is worthwhile to note that despite fitting the GCS model to three vantage point observations, the estimation of half-angle and aspect-ratio can still have considerable uncertainties. One way to reduce such uncertainty is to use observations from instruments that are placed away from the ecliptic, as reported by \cite{thernesien_2009}. So, in future, observations from the METIS \citep[][]{Fineschi_2012} on-board the Solar Orbiter \citep[][]{muller_2013} can be used to reach more precise estimation of these parameters. For the five CMEs, we found that the face-on width starts in the range of 10-30$^\circ$ which expands and becomes constant at 60-90$^\circ$. 
It should be noted that this was not possible in \cite{majumdar_2020}, since only two vantage point observations were used (which is also the case for the fifth event in Table~\ref{table} in this work), which often leads to a degeneracy in the $\alpha$ and $\gamma$ parameters (as reported in \cite{majumdar_2020,thernesien_2009}. Thus showing the importance of studying the true width of a CME, rather than the projected width, as the later is highly dependent on the observer's line of sight (LOS). It must be noted that many of the earlier studies have ignored LOS effects in the CME width, and hence statistical studies on the width distribution \citep[such as][and references therein]{pant_2021} can suffer from these projection effects.

%from the Atmospheric Imaging Assembly (AIA) of the Solar Dynamics Observatory \citep[SDO;][]{aia}  and STEREO/EUVI with SOHO/LASCO and STEREO/COR--2

%\begin{figure*}[h]
%\gridline{\fig{cme_kcor_20140212.png}{0.45\textwidth}{(a)}
   %       \fig{cor1a_20140212.png}{0.45\textwidth}{(b)}
        % }
%\gridline{\fig{cor1b_20140212.png}{0.45\textwidth}{(c)}
 %        \fig{kcor_20140212.png}{0.45\textwidth}{(d)}
  %       }
%\caption{The fitting of the GCS flux-rope to the pair of COR-1 images and K-Cor image.}
%\label{fig1}
%\end{figure*}

\subsection{Evolution of modelled CME volume}

\begin{figure}
    \centering
    \gridline{\fig{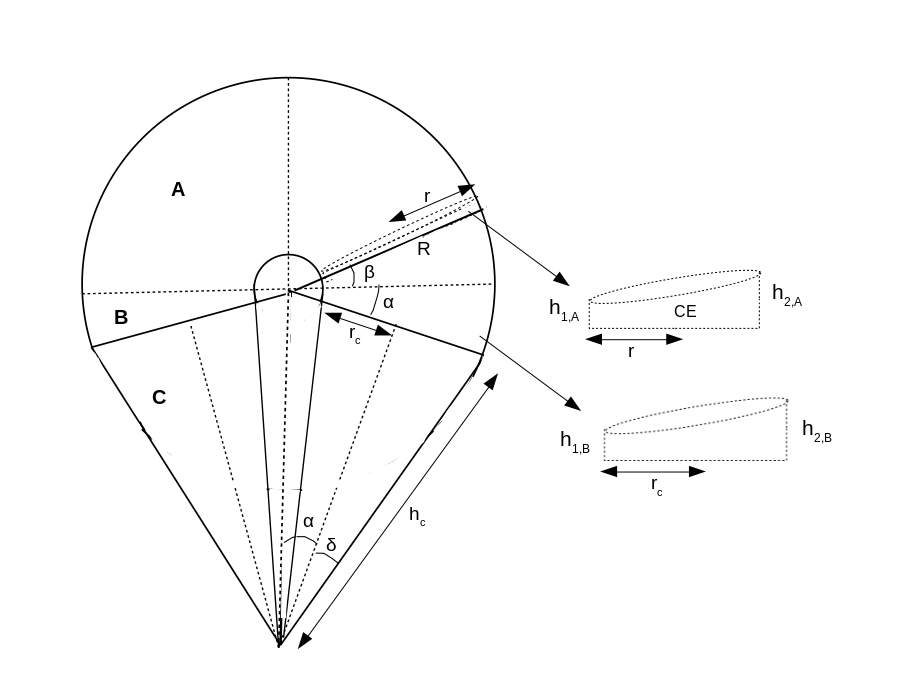}{0.8\textwidth}{}}
    \caption{A schematic for the estimation of CME volume from the GCS reconstructed 3D flux--rope structure. The entire CME volume can be sub divided into three parts, A: the ellipsoidal front, B: the asymmetric middle disc and C: the conical legs. The figure is adapted from \cite{Holzknecht_2018}}
    \label{gcs_vol}
\end{figure}

As reported by \cite{Holzknecht_2018}, the volume of a CME can be estimated from the GCS model, and it can be considered to be comprised of three parts, an ellipsoidal leading front (A in Figure~\ref{gcs_vol}), a middle asymmetric disc (B in Figure~\ref{gcs_vol}) and the conical legs (C in Figure~\ref{gcs_vol}). It should be noted here that all mathematical expressions are based on the work of \cite{Holzknecht_2018}. In order to calculate the volume of part A (V$_{\mathrm{A}}$), it is assumed that the ellipsoidal front is made of very thin asymmetric cylindrical elements (CE in Figure~\ref{gcs_vol}). Now, the entire volume of A is covered with  the angle $\beta$  (see Figure~\ref{gcs_vol}) ranging from $0^{\circ}$ to $90^{\circ}$. So, we divide $\beta$ into small fractions $\gamma$, and thus each CE consists of a constant $\gamma$,

\begin{equation}
    \gamma \, = \, \beta/n
\end{equation}

where $n$ is the number of thin CEs. It should be noted that each of these cylindrical elements (CEs) have two different heights h$_{1,A}$ and h$_{2,A}$ where the second height is greater than the first height (see Figure~\ref{gcs_vol} right panel). These heights are as follows :

\begin{equation}
   \mathrm{h_{2,A}} \, = \, \mathrm{R\,tan (\gamma)}
\end{equation}
 and,
 
 \begin{equation}
      \mathrm{h_{1,A}} \, = \, \mathrm{(R-2r)\,tan (\gamma)}
 \end{equation}

Using these, the volume of each of these thin elements can be estimated as,

\begin{equation}
    \mathrm{V_{CE}} \, = \, \mathrm{\pi r^2 \left( \frac{ h_{1,A}\,+h_{2,A}}{2} \right) } \, = \, \mathrm{\pi r^2(R-r)tan(\gamma)}
\end{equation}

So, summing over all these elements, V$_{\mathrm{A}}$ is calculated as follows :

\begin{equation}
    \mathrm{V_A} \, = \, \sum_{\gamma} \mathrm{\pi r^2(R-r)tan(\gamma)}
\end{equation}

Similarly, volume of part B (V$_{\mathrm{B}}$) is calculated for a cylinder with different heights h$_\mathrm{1,B}$ and h$_\mathrm{2,B}$. From Figure~\ref{gcs_vol}, let

\begin{equation}
    \mathrm{r|_{\beta = 0}} \, = \, \mathrm{r_0}
\end{equation}
and,

\begin{equation}
    \mathrm{R|_{\beta = 0}} \, = \, \mathrm{R_0}
\end{equation}

With these, we get the two heights of the cylinder as,

\begin{equation}
    \mathrm{h_{2,B}} \, = \, \mathrm{R_0\,Sin (\alpha)}
\end{equation}

and,

\begin{equation}
    \mathrm{h_{1,B}} \, = \, \mathrm{(R_0-2r_0)\,Sin (\alpha)}
\end{equation}

which gives the volume as 

\begin{equation}
    \mathrm{V_B} \, =  \, \mathrm{\pi r_c^2 \left( \frac{ h_{1,B}\,+h_{2,B}}{2} \right) } \, = \, \mathrm{\pi r_c^2(R_0-r_0)Sin(\alpha)}
\end{equation}

where, $\mathrm{r_c}$ from Figure~\ref{gcs_vol} and from \cite{thernesien_2011} is

\begin{equation}
    \mathrm{r_c} \, = \, \mathrm{h_c}Sin(\delta) \, = \, \mathrm{kh_c}
\end{equation}
where $\mathrm{h_c}$ is the length of the conical legs. Finally, for the third part (C), which are the legs of the CME, it is simply the volume of the cone, which is the following:

\begin{equation}
    \mathrm{V_C} \, = \, \mathrm{\frac{1}{3}\,\pi\,r_{c}^{2}\,h_c}
\end{equation}

where, $\mathrm{h_c}$ from \cite{thernesien_2009,thernesien_2011} is related to the GCS parameters as:

\begin{equation}
    \mathrm{h_{front}} = \mathrm{h_c}\,\frac{1}{1-\mathrm{k}}\,\frac{1+\mathrm{sin(\alpha)}}{\mathrm{cos(\alpha)}}
\end{equation}

Now,  R and r are a function of the GCS model parameters k (aspect-ratio), h (height) and $\alpha$ (half-angle) and can be found from \cite{thernesien_2011}. Since the model is axisymetric, the total volume will thus be:

\begin{equation}
    \mathrm{V_T} \, = \, 2(\mathrm{V_A\, + \, V_B\, + \, V_C})
\end{equation}

 Thus using the above three GCS model parameters, the modelled volume of the CME can be studied. A study of the GCS volume evolution was reported earlier by \cite{Holzknecht_2018}, but they studied the volume evolution in the greater heights (15 - 215 R$_{\odot}$). \cite{Temmer_2021} also used the GCS volume to study the CME density evolution with height in the outer corona (in the height range 15 - 30 R$_\odot$). But, in these studies, the crucial information of the volume evolution in the inner corona was missing.  It should be noted that although CMEs are known to evolve self-similarly in the outer corona \citep[see][]{Subramanian_2014}, yet in the inner corona, their propagation is non self-similar \citep[refer][]{cremades_2020,majumdar_2020}. Thus, a study of the evolution of modelled CME volume in the inner corona demands our attention.  In this regard, although \cite{Temmer_2021} used the GCS volume to estimate the densities of the magnetic ejecta and the sheath regions, yet an understanding of the volume evolution of the CME leading front and the CME legs have somehow evaded our understanding. With the incorporation of K-Cor observations with COR-1 for GCS reconstruction, we now address these limitations in our understanding of CME volume evolution. Thus, we study the evolution of modelled volume of the CMEs in 3D in the inner and outer corona, separately for the different sections of the CME volume (A - the ellipsoidal front, B - the asymmetric disc and C - the conical legs as shown in Figure~\ref{gcs_vol}). 

\begin{figure}
\gridline{\fig{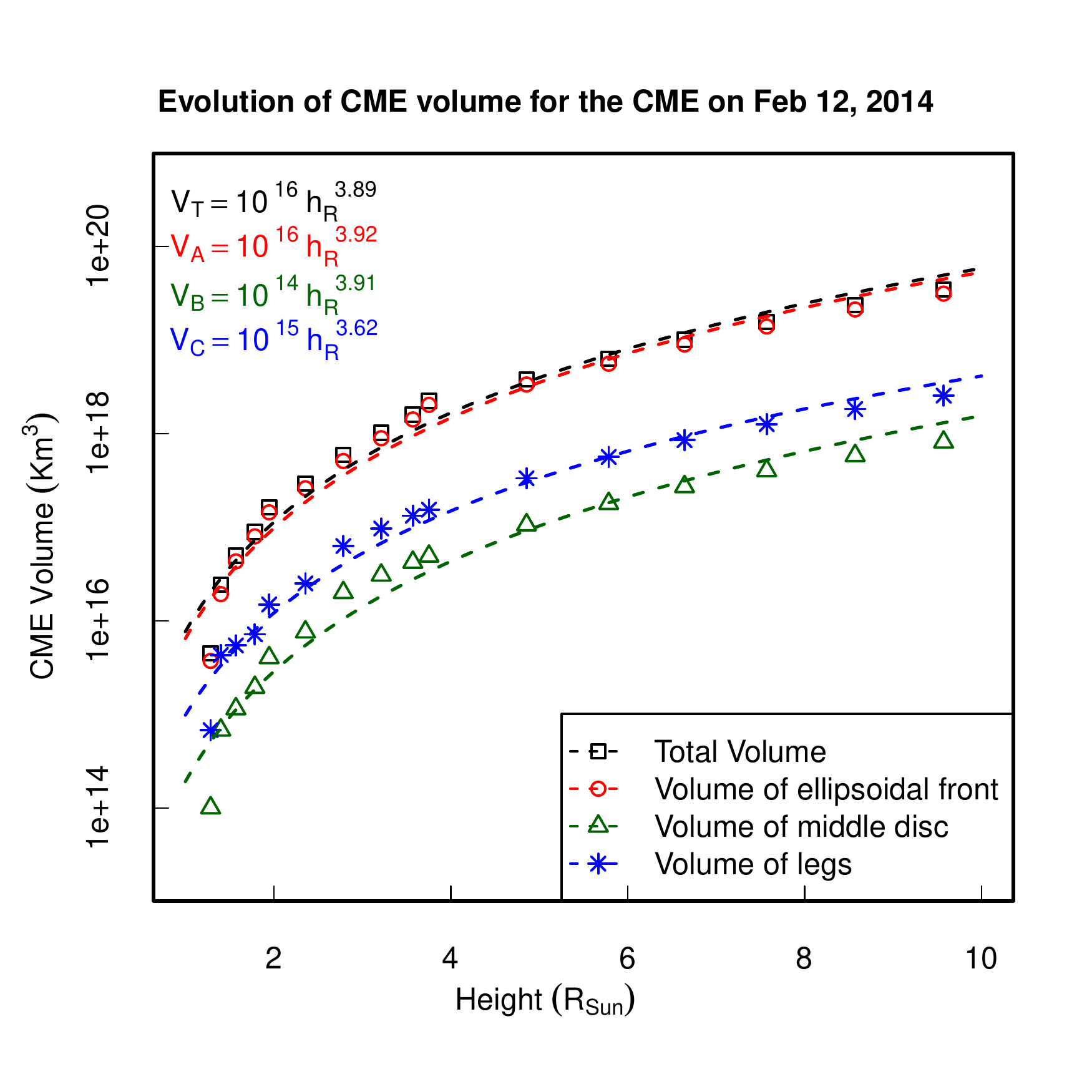}{0.33\textwidth}{(a) }
          \fig{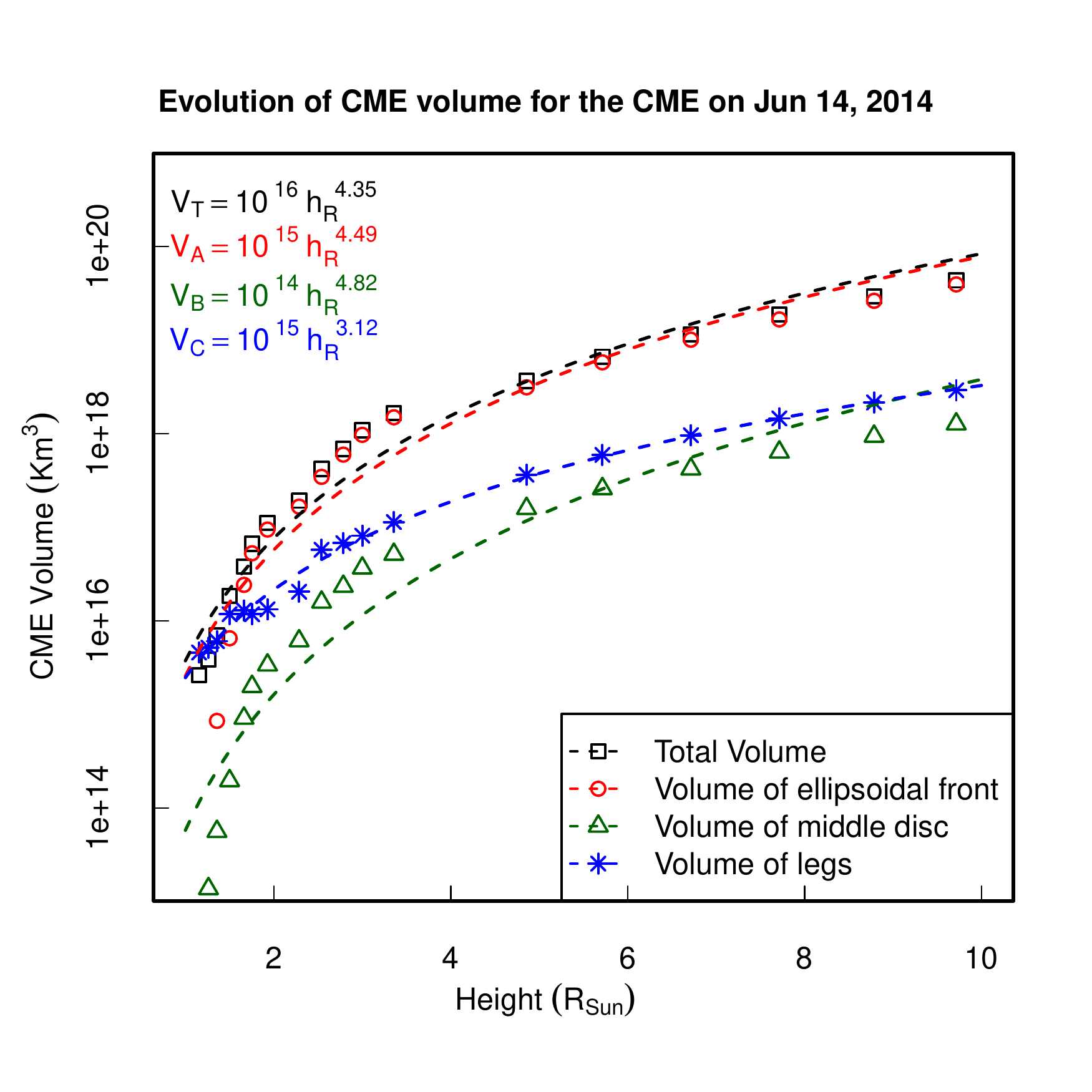}{0.33\textwidth}{(b) }
        \fig{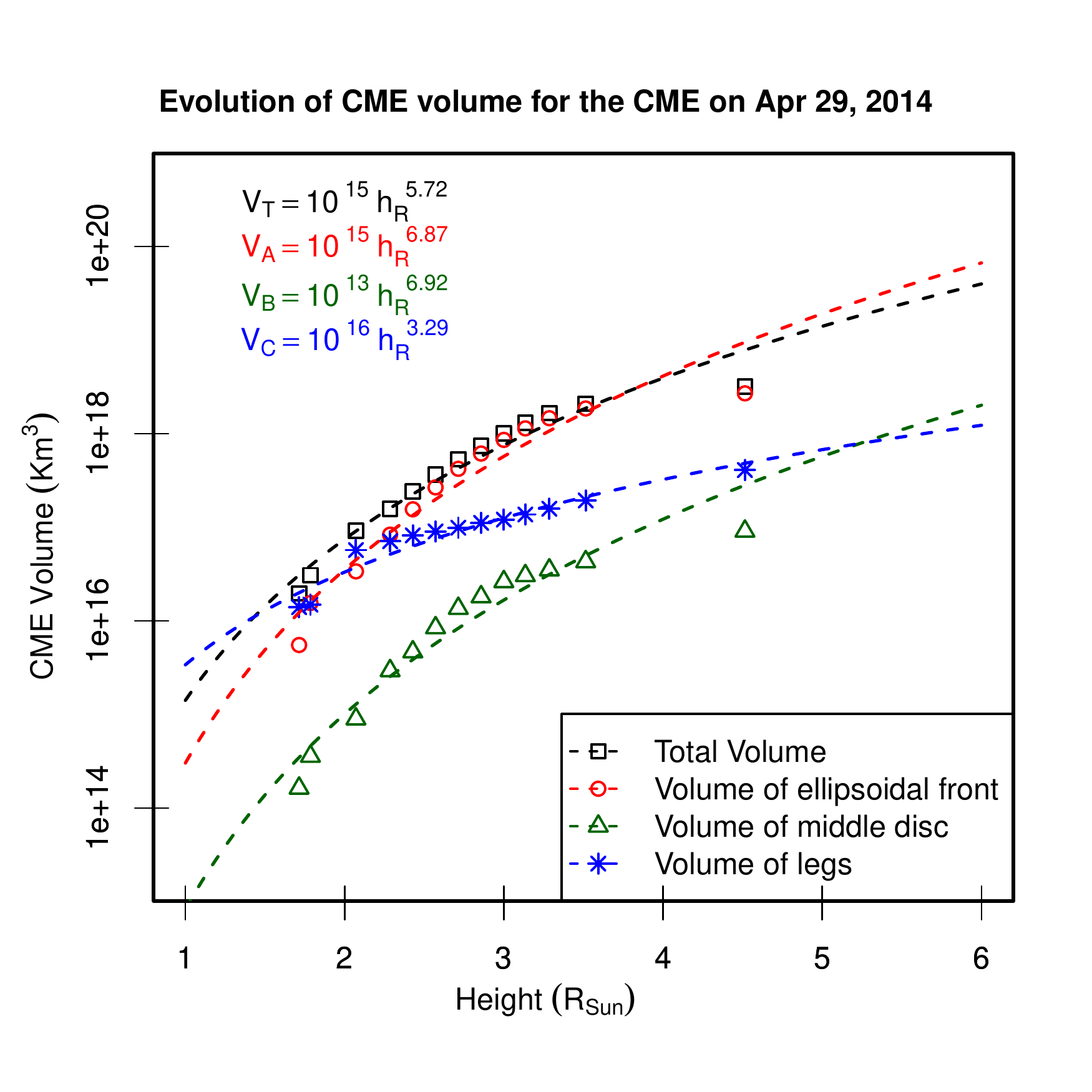}{0.33\textwidth}{(c) }
          } 
\gridline{\fig{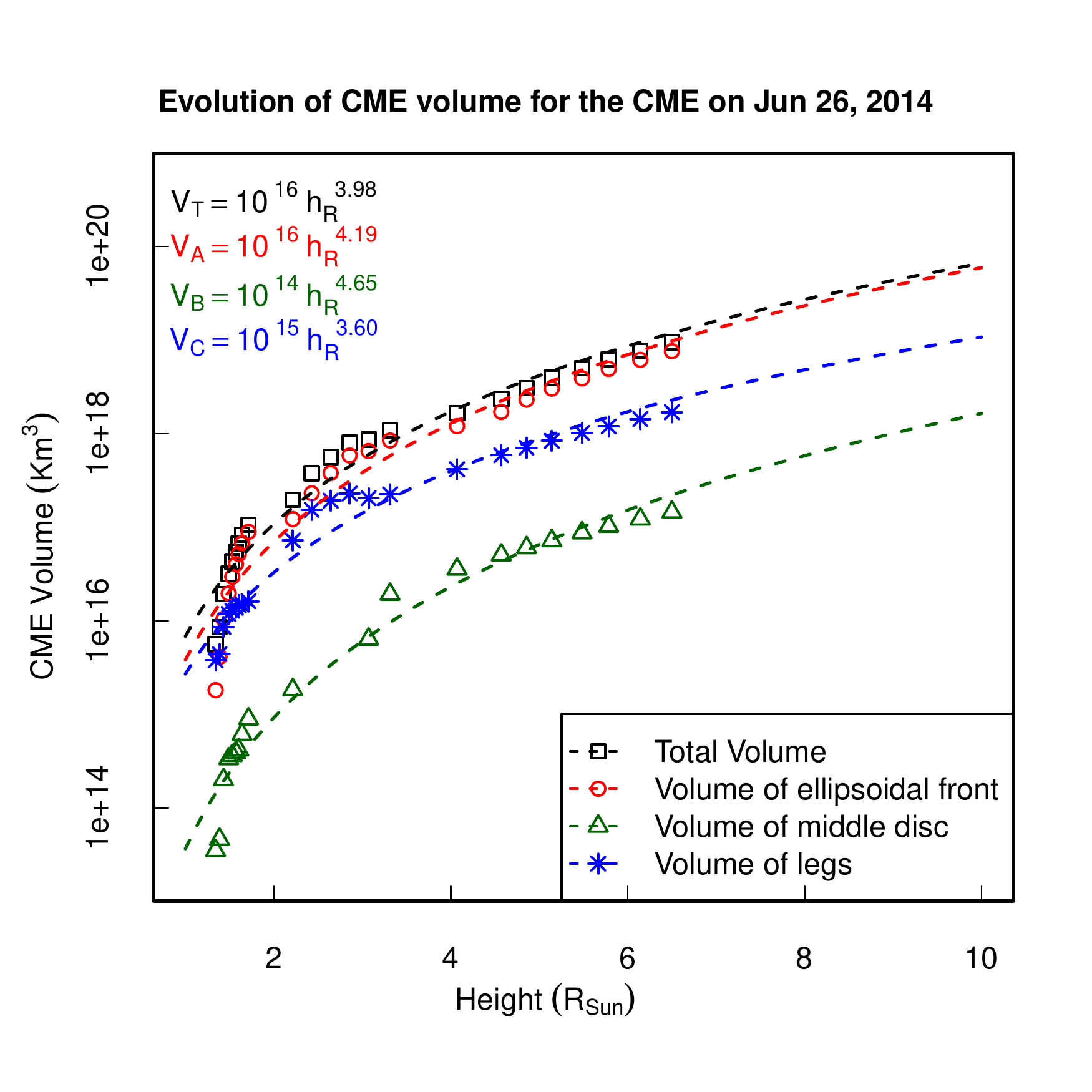}{0.33\textwidth}{(d) }
          \fig{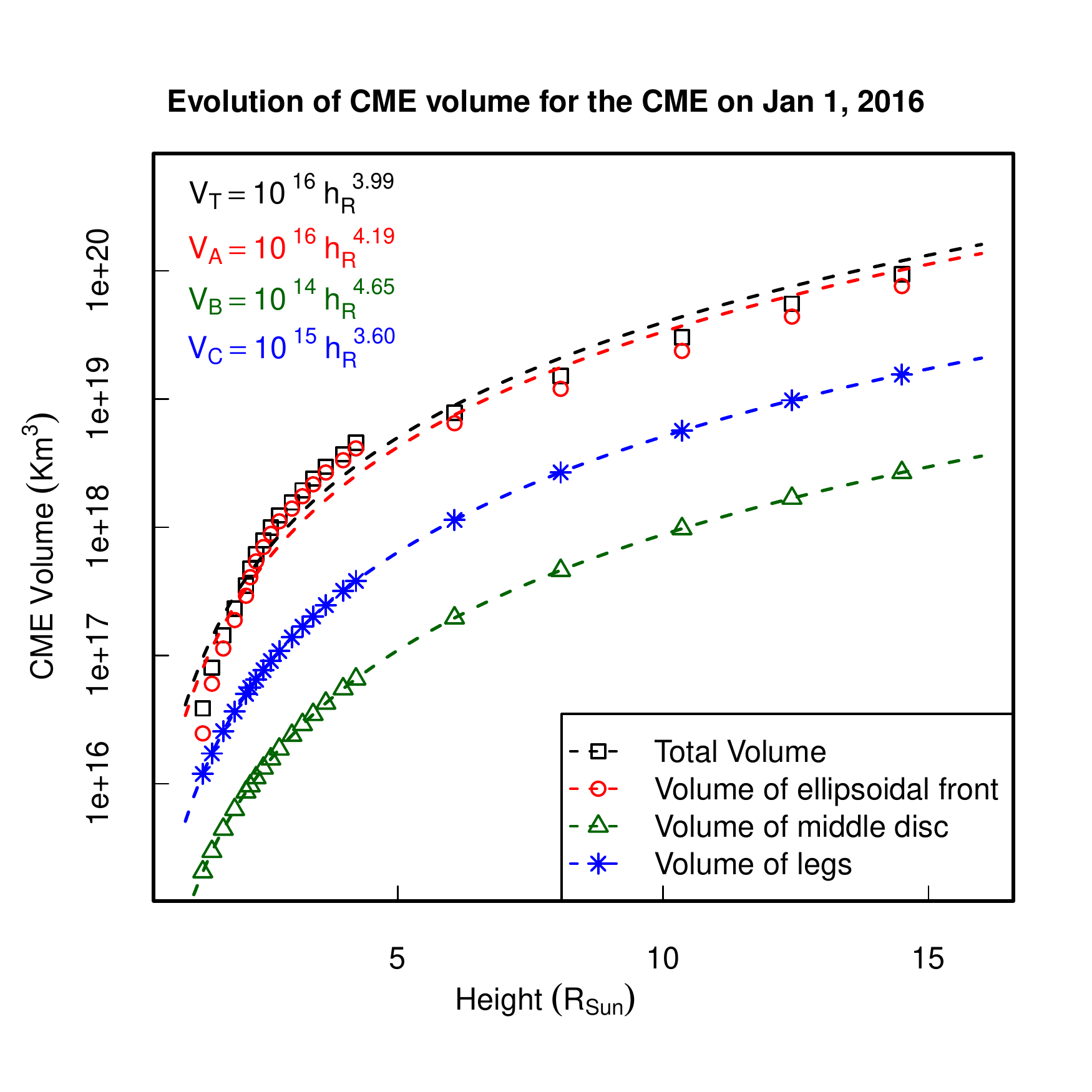}{0.33\textwidth}{(e) }
          }           
\caption{The evolution of the modelled CME volume and its different parts. The color coded plots denote evolution of different volume elements (the ellipsoidal front, the middle asymmetric disc and the conical legs) in the inner and outer corona.}
    \label{vol_profiles}
\end{figure}

 In Figure~\ref{vol_profiles}, we plot the modelled total volume (V$_\mathrm{T}$ in black) evolution of the five CMEs with the distance from the Sun in (a), (b), (c), (d) and (e). We then fit a power law that reflects the dependence of CME volume on the distance from the Sun, as the CME propagates outwards. This is the first time that any power law relation is reported for the evolution of modelled CME volume with height. Also since we have the volume estimated separately for the ellipsoidal front, the middle asymmetric disc and the conical legs of the CME, we study the evolution of these volumes as well and fit a power law to them for a better understanding.  For instance, it is in the inner corona where the CME starts forming, and thus studying the volume evolution of different parts of CMEs will enlighten us on the fact that whether CMEs retain their shape as they propagate from the inner to the outer corona. Further, a study of the associated power law profiles will help us understand the scale free behaviour of the volume expansion of CMEs with height. In other words, a single power law for all the different parts of the CME volume would imply a single unified mechanism that drives the volume expansion of CMEs, while different power laws would imply a differential volume expansion, and hence the possibility of different driving mechanisms. In addition to that, if the mechanism of acceleration and expansion of CME (that in turn affects the volume) is the same in inner and outer corona, then a single power law should be followed by the volume evolution profile in the inner and outer corona. However, if the power laws are different in the inner and outer corona, then that would imply that probably the mechanism of increase in volume might be different in inner and outer corona (as an outcome of the Lorentz force in inner corona and pressure difference in outer corona). Thus all these possibilities motivated us to probe the evolution of modelled CME volume in the inner and outer corona. 

\begin{center}
\begin{table}[h]
    \centering
    \begin{tabular}{c|c|c|c|c}
    \hline \hline
        Date & Volume Segment & Empirical Relation & R$^2$ Values & P - Values \\
        \hline \hline
      \multirow{4}{*}{2014 February 12} & Total (T) & $\mathrm{V_T\,=\,10^{16}\:h_R^{3.89}}$ & 0.96 & $1.5\,\times10^{-11}$ \\
       & A & $\mathrm{V_A\,=\,10^{16}\:h_R^{3.92}}$ & 0.96 & $2.0\,\times10^{-11}$ \\
        & B & $\mathrm{V_B\,=\,10^{14}\:h_R^{3.91}}$ & 0.96 & $4.7\,\times10^{-11}$ \\
         & C & $\mathrm{V_C\,=\,10^{15}\:h_R^{3.62}}$ & 0.97 & $2.9\,\times10^{-12}$ \\
         \hline
        \multirow{4}{*}{2014 June 14} & Total (T) & $\mathrm{V_T\,=\,10^{16}\:h_R^{4.35}}$ & 0.96 & $8.3\,\times10^{-13}$ \\
       & A & $\mathrm{V_A\,=\,10^{15}\:h_R^{4.49}}$ & 0.92 & $5.1\,\times10^{-9}$ \\
        & B & $\mathrm{V_B\,=\,10^{14}\:h_R^{4.82}}$ & 0.91 & $4.1\,\times10^{-9}$ \\
         & C & $\mathrm{V_C\,=\,10^{15}\:h_R^{3.12}}$ & 0.99 & $2.2\,\times10^{-16}$ \\ 
         \hline
        \multirow{4}{*}{2014 June 26} & Total (T) & $\mathrm{V_T\,=\,10^{16}\:h_R^{3.98}}$ & 0.95 & $2.3\,\times10^{-15}$ \\
       & A & $\mathrm{V_A\,=\,10^{16}\:h_R^{4.19}}$ & 0.93 & $1.9\,\times10^{-13}$ \\
        & B & $\mathrm{V_B\,=\,10^{14}\:h_R^{4.65}}$ & 0.96 & $2.7\,\times10^{-14}$ \\
         & C & $\mathrm{V_C\,=\,10^{15}\:h_R^{3.60}}$ & 0.96 & $2.2\,\times10^{-16}$ \\ \hline
        \multirow{4}{*}{2014 April 29} & Total (T) & $\mathrm{V_T\,=\,10^{15}\:h_R^{5.72}}$ & 0.95 & $1.2\,\times10^{-8}$ \\
       & A & $\mathrm{V_A\,=\,10^{15}\:h_R^{6.87}}$ & 0.93 & $1.0\,\times10^{-7}$ \\
        & B & $\mathrm{V_B\,=\,10^{13}\:h_R^{6.92}}$ & 0.94 & $7.3\,\times10^{-8}$ \\
         & C & $\mathrm{V_C\,=\,10^{16}\:h_R^{3.29}}$ & 0.93 & $1.2\,\times10^{-7}$ \\ \hline \hline 
          \multirow{4}{*}{2016 January 1} & Total (T) & $\mathrm{V_T\,=\,10^{16}\:h_R^{3.99}}$ & 0.96 & $2.8\,\times10^{-15}$ \\
       & A & $\mathrm{V_A\,=\,10^{16}\:h_R^{4.19}}$ & 0.95 & $9.3\,\times10^{-14}$ \\
        & B & $\mathrm{V_B\,=\,10^{14}\:h_R^{4.65}}$ & 0.99 & $2.2\,\times10^{-16}$ \\
         & C & $\mathrm{V_C\,=\,10^{15}\:h_R^{3.60}}$ & 0.99 & $2.2\,\times10^{-16}$ \\ \hline \hline 
    \end{tabular}
    \caption{The empirical relations for the volume evolution of CMEs with the corresponding R$^2$ values and P - values for the different sections of the CME.}
    \label{fit_values}
\end{table}  
\end{center}

The details of the fitted power laws are given in Table~\ref{fit_values}. In order to appreciate the fitted empirical relations, we provide the associated R$^2$ values that shows how well our model succeeds in determining the strength of the relationship between our model and the dependent variable on a 0 - 1 scale. We also provide the associated p-value which shows the statistical significance of the fitted model. The average significance level was found to be $0.05$ on average, and thus models with p-values lesser than $0.05$ implies statistically significant result. We find that the power law index for the total volume ranges between 3.89 - 5.72, thus indicating that the volume of a CME keeps increasing with distance from the Sun within the investigated height. We further find that the volume of the leading ellipsoidal front (V$_\mathrm{A}$) and that of the middle disc (V$_\mathrm{B}$) varies with a higher power law index (ranging between 3.92 - 6.87 and 3.91 - 6.92 respectively) than that of the total volume, while the volume of the conical legs (V$_\mathrm{C}$) varies with a much lower power law index (ranging between 3.12 - 3.62), thus indicating a differential volume evolution throughout a CME. This once again reflects the significance of studying both the FO and EO widths of a CME. It is important to note that the volume of the legs of the CME is largely influenced by the EO width of the CME, while the volume of the other two sections are influenced by both the FO and EO widths. However, it must also be kept in mind that the estimation of the volume of the legs by this method is possible only for CMEs with small aspect ratios (as is the case for majority of the events studied, please see Table~\ref{table}), which will enable the identification of two separate legs distinctly (as seen in the K-Cor images in Figure~\ref{gcs_fits}). For future studies on CMEs with large aspect ratios, it should be kept in mind that there will be a substantial overlap of the legs and hence the estimation of the volume of the legs might be misleading in such cases. 

 From Figure~\ref{width_profile}, it can be seen that the EO widths are much lesser in magnitude as compared to the FO widths (which is an expected outcome of the geometry of the GCS model), and this is further reflected in the power laws as a slower increase of the volume of the legs of the CME as compared to the ellipsoidal front and middle disc. We also note that the power law for the total volume is substantially greater for the CME on 2014 June 14 and 2014 April 29, as compared to the other three cases. We found that these two CMEs were ejected from erupting quiescent prominences, while the other three events were ejected from active regions. Recently, \cite{pant_2021} have reported a higher power law index for the width distribution of CMEs connected to quiescent erupting prominences than those connected to active regions. It seems the volume of a CME too shows a similar imprint of the source region, but our conclusion in this work is based on only five events and hence an extension of this study to a much larger sample set of events will help in better establishing our conclusions. In future these results will also provide better inputs to study the dynamics of mass accretion by the CMEs as they evolve in the lower heights.

\begin{center}
\begin{table}[h]
    \centering
    \begin{tabular}{c|c|c|c|c}
    \hline \hline
        Date & Region & Empirical Relation & R$^2$ Values & P - Values \\
        \hline \hline
      \multirow{2}{*}{2014 February 12} & $\mathrm{h_R}\,<\,4\mathrm{R_{\odot}}$ & $\mathrm{V_{}\,=\,10^{16}\:h_R^{4.93}}$ & 0.95 & $1.5\,\times10^{-6}$ \\
       & $\mathrm{h_R}\,>\,4\mathrm{R_{\odot}}$ & $\mathrm{V_{}\,=\,10^{16}\:h_R^{3.28}}$ & 0.99 & $2.3\,\times10^{-7}$ \\
         \hline
        \multirow{2}{*}{2014 June 14} & $\mathrm{h_R}\,<\,4\mathrm{R_{\odot}}$ & $\mathrm{V_{}\,=\,10^{15}\:h_R^{6.15}}$ & 0.98 & $1.5\,\times10^{-10}$ \\
       & $\mathrm{h_R}\,>\,4\mathrm{R_{\odot}}$ & $\mathrm{V_{}\,=\,10^{16}\:h_R^{3.54}}$ & 0.99 & $2.5\,\times10^{-8}$ \\
         \hline
        \multirow{2}{*}{2014 June 26} & $\mathrm{h_R}\,<\,4\mathrm{R_{\odot}}$ & $\mathrm{V_{}\,=\,10^{16}\:h_R^{5.08}}$ & 0.91 & $2.4\,\times10^{-8}$ \\
       & $\mathrm{h_R}\,>\,4\mathrm{R_{\odot}}$ & $\mathrm{V_{}\,=\,10^{16}\:h_R^{3.82}}$ & 0.99 & $3.1\,\times10^{-9}$ \\
        \hline
        \multirow{2}{*}{2014 April 29} & $\mathrm{h_R}\,<\,4\mathrm{R_{\odot}}$ & $\mathrm{V_{}\,=\,10^{15}\:h_R^{5.72}}$ & 0.95 & $1.2\,\times10^{-8}$ \\
       & $\mathrm{h_R}\,>\,4\mathrm{R_{\odot}}$ & -- & -- & -- \\
        \hline
        \multirow{2}{*}{2016 January 1} & $\mathrm{h_R}\,<\,4\mathrm{R_{\odot}}$ & $\mathrm{V_{}\,=\,10^{16}\:h_R^{4.17}}$ & 0.99 & $3.6\,\times10^{-14}$ \\
       & $\mathrm{h_R}\,>\,4\mathrm{R_{\odot}}$ & $\mathrm{V_{}\,=\,10^{17}\:h_R^{2.47}}$ & 0.97 & $0.0002$ \\
        \hline \hline 
    \end{tabular}
    \caption{The empirical relations for the volume evolution of CMEs in the inner and outer corona with the corresponding R$^2$ values and P - values.}
    \label{fit_values2}
\end{table}  
\end{center}

\begin{figure}
\gridline{\fig{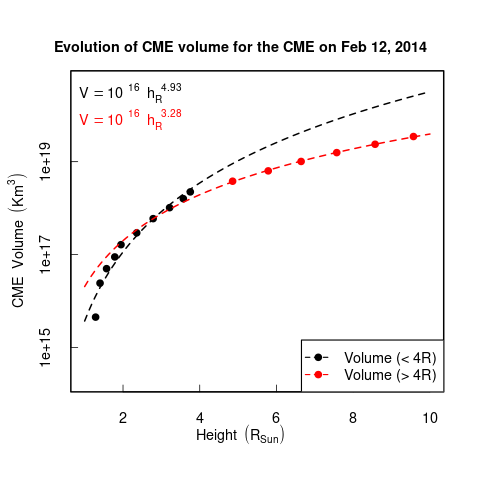}{0.45\textwidth}{(a) }
          \fig{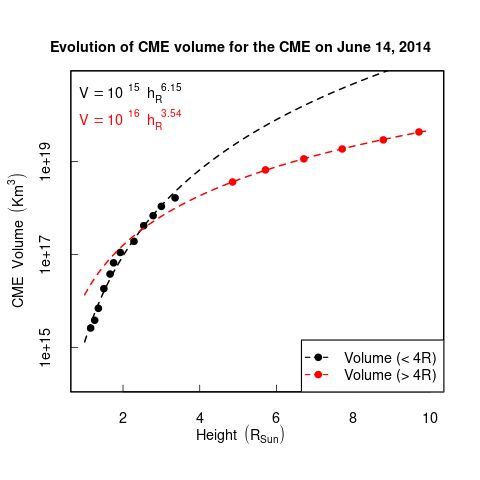}{0.45\textwidth}{(b) }
          } 
\gridline{\fig{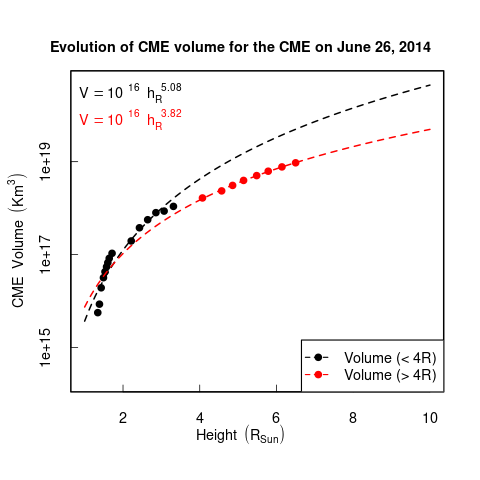}{0.45\textwidth}{(c) }
          \fig{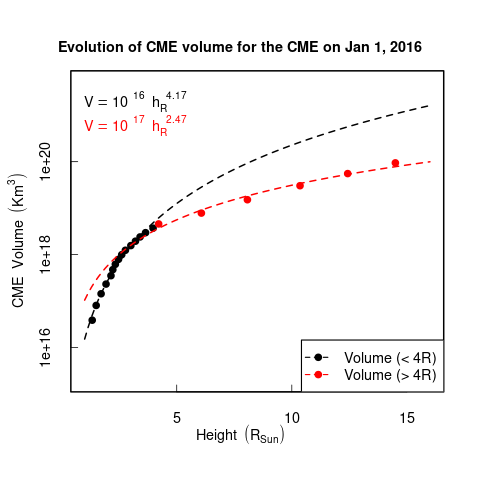}{0.45\textwidth}{(d) }
          }           
\caption{The evolution of the total modelled CME volume in the inner and outer corona. The data points and curve in black mark the volume in inner corona ($<\,4$R$_{\odot}$), while the ones in red are for those in the outer corona ($>\,4$R$_{\odot}$).}
    \label{vol_in_out}
\end{figure}

As discussed earlier, CMEs tend to evolve self-similarly in the outer corona, while the evolution in the inner corona is non self-similar. This change in behaviour of the CMEs provoked us further to study the total volume evolution of CMEs separately in the inner and outer corona. A close look at Figure~\ref{vol_profiles} hints  that the total volume of the CMEs shows different characteristics at different heights and that a single power law is not able to fit the volume evolution for the entire height range. So in Figure~\ref{vol_in_out}, we plot the evolution of the modelled total volume of the CMEs, and we fit two separate power law profiles for the evolution of volume below and beyond $4\,\mathrm{R_{\odot}}$ (please see Table~\ref{fit_values2} for the details of fitting). It should be noted here that for the event on 2014 April 29, we could not track the CME much further in the COR-2 FOV, as the leading edge got depleted, and was difficult to track. Thus, it was not possible to study the evolution of the modelled volume in the outer corona. We find that the volume evolution for all the events follow different power law profiles in the inner and outer corona. We find that the volume increases much more rapidly in the lower heights in the inner corona, as compared to the outer corona, and thus this clearly indicates towards the possibility of two different expansion mechanisms for CMEs in these two height regimes. The initial rapid expansion of volume can be attributed to the rapid angular width expansion in the inner corona as was recently reported by \cite{cremades_2020} and \cite{majumdar_2020}, while it seems the relatively slower volume expansion of CMEs in the outer corona might be a consequence of the total pressure difference in the inside and outside of the CME. These results thus strongly indicate how the kinematic properties of CMEs in the inner corona are strikingly different from the properties in the outer corona, lending support to the recent report by \cite{majumdar_2021b}. It is also worthwhile to note the significance of the inclusion of K-Cor data along with the COR-1 data in order to arrive at these results. The measurements in the K-Cor FOV have facilitated in distinctly distinguishing the contrast in the evolution of modelled CME volume in the inner and outer corona.  

%\subsection{Limitations and scope for future}

\section{Discussions and Conclusions} \label{sec4}

We  first present the feasibility of implementing GCS on the K-Cor data-sets for the first time thereby providing additional vantage point for 3D reconstruction of CMEs in the inner corona.  A proof of concept of this application is presented in Figure~\ref{gcs_fits} by fitting the GCS model to the near-simultaneous images of K-Cor along with the observations from STEREO/SECCHI coronagraphs. The combined coronagraphic observations of K-Cor and STEREO/COR-1 in the inner corona and STEREO/COR-2 and SoHO/LASCO in the outer corona, allowed us to track and study the true evolution of CMEs in white-light, covering a FOV starting from as low as 1.1 R$_\odot$ which was never achieved earlier. This was possible once the GCS parameters for the CMEs were fixed by the three above-mentioned vantage points which were back-traced in the K-Cor FOV to heights of $\approx$1.1 R$_\odot$. This facilitated in capturing the initial impulsive phase of the CMEs, where the kinematic parameters are known to change rapidly.

We were able to track the initial rapid expansion of CMEs in these lower heights, and thanks to three vantage point observations, we found that CMEs expand along both the axial and lateral directions rapidly in the initial part of the trajectory till a height of 3 R$_{\odot}$, after which it saturates to a constant value. It should be noted here that for the CME in 2016 only two vantage points were available, while for the heights below the COR-1 FOV, only K-Cor observations were used.
We noted that the CMEs can expand from $\sim$10$^\circ$ to more than 90$^\circ$ in face-on width within inner corona. For the sample of CMEs we fitted, it could be identified that even though there was not much impulsiveness in the radial kinematics of the CMEs in the inner corona, we see a considerable expansion in their widths. An extension of this study on a larger data-set will  provide better understanding of the Lorentz force in early kinematics of CMEs. 
In the future, an estimation of the true acceleration duration and magnitude can also be done at lower heights without any underestimate in the mentioned quantities, which was not possible in \cite{majumdar_2020, cremades_2020}. It is worthwhile to note that we were able to do this using only white-light data (within the limitations of the GCS model), hence ensuring that any ambiguity arising from tracking a CME in EUV and white-light is further evaded. Thus, this work will largely help in improving on the shortcomings in previous studies on CME kinematics \citep{bein_2011, Subramanian_2014, cremades_2020, majumdar_2020}. We further used the GCS model geometry to estimate the modelled total volume of the CME and also the modelled volumes of the ellipsoidal leading front, the asymmetric disc in middle and the conical legs separately. It should be noted that a correct estimation of the volumes of the different segments of the flux-rope requires the unambiguous identification of the inner edge of the flux-rope. But the identification of the inner edge of flux-rope is very difficult and tricky in the coronagraph images, and even if identified, it will suffer from high observer bias. However, provided the FOV of the coronagraph provides observation at the absolute lower heights to face-on CMEs (as is the case for K-Cor images in Figure~\ref{gcs_fits}), the inner edge of the flux-rope can be identified and gauged at the CME legs. Here, in this context, all CMEs analyzed in this work are assumed to be oriented face-on. We report for the first time, a power law variation of the modelled CME volume with distance from the Sun. We also found that the power law is higher for the ellipsoidal front and the disc than that for the conical legs, thus indicating that the volume expansion is dominated by the former two parts while the volume of the legs increases relatively slower, thus indicating that there is a differential volume expansion through a CME as it propagates from the inner to the outer corona. In this context, it must also be kept in mind that the estimation of the volume of the legs by this method is possible only for CMEs with small aspect ratios (as is in our case, please see Table~\ref{table}), which will enable the identification of two separate legs distinctly (as seen in the K-Cor images in Figure~\ref{gcs_fits}). For future studies on CMEs with large aspect ratios, it should be kept in mind that there will be a substantial overlap of the legs and hence the estimation of the volume of the legs might be misleading in such cases. We also studied the evolution of the modelled total volume of the CMEs in the inner and outer corona, and we found that CMEs tend to follow two distinctly different power law profiles below and beyond $4\,\mathrm{{R_{\odot}}}$. This indicates at the possibility of two different expansion mechanisms of CMEs in the inner and outer corona. We believe these results need further attention in the future which will help us better understand the coupling of CME kinematics as they evolve from the inner to the outer corona.It is worthwhile to note that as a consequence of the constraints in the fitting procedure in the absolute lower heights (as outlined in Section~\ref{fit_algo}), the height measurements will get influenced for CMEs which get deflected in the lower heights. Now, although the CMEs studied in this work did not show any appreciable deflection, yet in future such considerations should be kept in mind while studying CMEs that get deflected, as that will increase the uncertainty in the height measurements. In addition to that, this work ignores rotation of the CME in the lower heights as no such observable evidence was noted. Now despite the fact that no such observable signatures of deflections and rotations were noticed, it is worth noting that it is not that trivial to conclude on these properties, solely based on visual inspections. Hence, in future, possibly with the inclusion of above the ecliptic data from METIS on-board the Solar Orbiter, or observations from missions placed at the L5 point, we will be able to arrive at much stronger and better constrained conclusions. Thus in future for CMEs that exhibit rotation, a change in the tilt--angle parameter should also be considered while estimating the volume of the CME. Also, consideration of these processes (rotation and deflection) in future studies will also help in better understanding the evolution of the volume of the front and legs of the CME.

In this context, it must also be noted here that these conclusions are specific to the geometry of the GCS model, which is an idealized geometrical figure that has its limitations and constrains \citep[see][]{thernesien_2009}. Regarding the evolution of the legs, the identification of two separate legs of the CMEs require observation at the absolute lower heights. Thus the legs can be identified in the K-Cor FOV, while its still not seen in the COR-1 FOV at the same time as shown in Figure~\ref{gcs_fits}, but it should also be noted that despite the promising FOV of K-Cor, yet the poor image quality due to challenges faced from being a ground based coronagraph makes it difficult to fit (refer the discussion in Section~\ref{event_selection}. In this regard, the upcoming ADITYA--L1 \citep{aditya-l1} mission with the Visible Emission Line Coronagraph \citep[VELC;][]{velc, Banerjee2017} (FOV : 1.05 -- 3 R$_{\odot}$) on-board and PROBA--3 \citep[][]{proba-3} (FOV : 1.02 - 3 R$_{\odot}$) with the giant Association de Satellites pour l'Imagerie et l'Interferométrie de la Couronne Solaire \citep[ASPIICS;][]{Lammy_2017} will provide much better data and hence will help in arriving at much stronger conclusion on the evolution of CME legs. Having said that, it must also be noted that a true estimation of the volume of CME legs require the CME to be seen face-on, as a face-on view will help in identifying the inner edges of the CMEs and hence the volume of their legs. The studied CMEs in this work are all seen face--on in the K-Cor FOV (please see Figure~\ref{gcs_fits}). Thus, in future, for a larger statistical study, the appearance of the CME ( being whether face--on or edge--on) should also be considered in the estimation of the volume of the CME legs. Apart from that, around one-third fraction of CMEs have been reported to have a flux-rope morphology \citep[see][]{Vourlidas_2013}, which happens to be the bedrock of the foundation of the GCS model, thus a study of these three separate sections of the flux-rope model of the CME will help us have a much better understanding on the validity of self-similar expansion, and thus provide more precise constraints to models that study flux-rope initiation and evolution

The cadence of K--Cor is better than that of COR--1, and this helped in tracking the CME more effectively in the lower heights, thus getting more number of data points in the impulsive phase. Since, the speed and acceleration of a CME are obtained by taking first and second order derivatives of the height--time data, it is essential to have as many number of data points as possible, especially in the initial impulsive phase, so that the derived quantities are better estimated \citep{byrne_2012}. In this regard, although K--Cor provides a cadence of fifteen seconds, but the signal to noise in that data is not good enough for confident tracking of the CMEs in most of the cases, which made us use the two minute cadence data. Now although this is a substantial improvement on the cadence of COR--1, yet it barely needs explanation that data with even better cadence will further aid in our understanding of this initial rapid impulsive phase of CMEs. For this, again the data from upcoming space missions like ADITYA--L1 with the VELC, and PROBA--3 with the ASPIICS will help in overcoming this limitation by providing high cadence data with good resolution. The significance of this extension of the GCS model also lies in the fact that, despite the loss of STEREO-B (and hence COR--1B data) from 2016, this modified GCS model will still enable stereoscopy in the inner corona for the 3D study of early kinematics of CMEs in white-light.

%\textcolor{cyan}{\bf Do CMEs Evolve Differently from Inner to Outer Corona?}
%\textcolor{cyan}{\bf On the Differential Evolution of CMEs in inner and outer corona}\\
%\textcolor{cyan}{\bf Combined Ground- and Space-based Stereoscopy Provides Evidence of Differential Evolution of CMEs!}

\begin{acknowledgments}

We thank the anonymous referee for the valuable comments that have improved the manuscript. The authors acknowledge Dipankar Banerjee for his unstinted support and motivation throughout this project. Courtesy of the Mauna Loa Solar Observatory, operated by the High Altitude Observatory, as part of the National Center for Atmospheric Research (NCAR). NCAR is supported by the National Science Foundation. The SECCHI data used here were produced by an international consortium of the Naval Research Laboratory (USA), Lockheed Martin Solar and Astrophysics Lab (USA), NASA Goddard Space Flight Center (USA), Rutherford Appleton Laboratory (UK), University of Birmingham (UK), Max-Planck-Institut for Solar System Research (Germany), Centre Spatiale de Li$\grave{e}$ge (Belgium), Institut d'Optique Th$\acute{e}$orique et Appliqu$\acute{e}$e (France), Institut d'Astrophysique Spatiale (France).We also acknowledge the SoHO team for LASCO data.
\end{acknowledgments}


\vspace{5mm}
\facilities{K-Cor/MLSO, COR-1+COR-2/STEREO-A/B, LASCO/SOHO}


\software{IDL, R \citep{R}}




%\bibliography{references_gcs_kcor}{}
\bibliographystyle{aasjournal}



\end{document}